\documentclass[11pt]{article}

\usepackage{authblk}
\usepackage{sectsty}
\usepackage{graphicx}
\usepackage{amsmath}
\usepackage{amssymb}
\usepackage{listings}
\usepackage{xcolor}
\usepackage{tikz}
\usepackage{siunitx}
\usepackage{enumitem}
\usepackage{longtable}
\usepackage{booktabs}
\usepackage[hidelinks]{hyperref}
\usepackage{cite}
\usepackage{url}
\usepackage[version=4]{mhchem}
\usepackage{wasysym}

\newlist{tasks}{itemize}{1}
\setlist[tasks]{label=$\square$}

\usetikzlibrary{arrows.meta}

\title{Exploring Initial CO\textsubscript{2} Transport Topologies for Germany's Carbon Management}
\author[1]{Toni Seibold}
\author[2,1]{Luna Lütz}
\author[1]{Tom Brown}
\affil[1]{Digital Transformation in Energy Systems, Technische Universität Berlin, Einsteinufer 25, 10587 Berlin, Germany}
\affil[2]{Fraunhofer Institute for Systems and Innovation Research (ISI), Breslauer Str. 48, 76139 Karlsruhe, Germany}

\date{}

\begin{document}
\maketitle

\begin{abstract}
    %Motivation
    \noindent Germany's climate target requires carbon capture and sequestration for residual emissions from hard-to-abate sectors such as cement production and waste incineration.
    Planning this infrastructure is challenging because capture investments and \ce{CO2} transport networks are strongly interdependent.
    Existing energy system models capture system-wide interactions but provide limited insight into robust transport topologies, while detailed infrastructure studies usually neglect feedbacks with the wider energy system.
    % Method
    This study combines graph-theoretic topology generation with a large-scale sector-coupled energy system model to evaluate alternative \ce{CO2} transport networks for Germany in 2035.
    We generate and evaluate 60 candidate topologies that differ in network length, sink accessibility and source prioritization.
    % Results
    The deployment of a domestic \ce{CO2} transport network reduces German consumer costs by around \SI{22}{bnEUR/a} relative to a scenario without \ce{CO2} pipelines.
    The resulting network is primarily used by industrial point sources, including process emissions, cement production and biomass-based carbon dioxide removal, while contributions from backup power generation remain comparatively small.
    Transport corridors are repeatedly selected and utilized in north-western Germany, reflecting the concentration of industrial \ce{CO2} sources and access to international sequestration routes.
    Early access to Dutch sink infrastructure provides particularly high system value.
    Compared to \SI{1500}{km}, as little as \SI{500}{km} of \ce{CO2} pipeline infrastructure captures most of the economic benefit when North Rhine-Westphalia is connected to the Netherlands, while also limiting long-term transport infrastructure lock-in.
    % Conclusion
    These findings suggest that the availability of \ce{CO2} transport infrastructure is more important than the exact topology once major industrial source regions and sink access points are connected.
    The proposed combination of graph-theoretic topology generation and integrated energy system modelling helps identify highly utilized transport corridors for early-stage \ce{CO2} infrastructure planning.
\end{abstract}

\noindent Keywords: Carbon capture and storage, \ce{CO2} transport infrastructure, Graph-based topology generation, Energy system modelling, Infrastructure planning

\section*{Highlights}
\begin{itemize}
    \item Graph-based topology generation enables evaluation of early German \ce{CO2} pipeline networks in a sector-coupled energy system model
    \item Industrial process emissions dominate \ce{CO2} capture and pipeline utilization
    \item \ce{CO2} infrastructure availability is more important than the exact route in 2035
    \item Under tight network budgets, early access to the Netherlands reduces long-term costs
\end{itemize}

% \section*{Graphical Abstract}

\section{Introduction}
\noindent\textbf{Motivation}

\noindent Germany has committed to achieving net-zero emissions by 2045.
To reach this ambitious goal, electrification and the deployment of green hydrogen are key pillars of Germany's decarbonization strategy.
However, certain emissions remain challenging to mitigate due to the absence of viable low-carbon alternatives.
To bridge the gap to net-zero and manage these hard-to-abate residual emissions carbon capture and sequestration (CCS) becomes an essential strategy.

Policy developments acknowledge this necessity.
The European Net-Zero Industry Act (NZIA) identifies CCS as vital for industrial decarbonization, while Germany's Carbon Management Law (KSpG) establishes a national legal framework specifically enabling CCS \cite{european_parliament_eu_2024, noauthor_kohlendioxid-speicherung-und-transport-gesetz_2025}.
Cement plants and waste incineration facilities are expected to be major contributors, but the framework also allows for other emissions to be permanently stored.
On the storage side depleted gas and oil fields in the North Sea offer significant sequestration potential.
Flagship projects such as \textit{Northern Lights}, \textit{Porthos} and the \textit{Delta Rhine Corridor} reflect the ambition and anticipated need for large-scale \ce{CO2} transport and sequestration \cite{european_union_northern_2026, european_union_delta_2025, european_union_co2_2026}.
Next to concrete projects, \textit{Open Grid Europe} is mapping a potential carbon backbone to connect emission sources with storage sites \cite{noauthor_co-netz_nodate}.
The major national energy system studies which inform policy makers regarding the energy transition in Germany acknowledge the future role of CCS in Germany and find a necessary volume between \SIrange{11.5}{70}{Mt/a} for sequestration but do not investigate the corresponding infrastructure needs \cite{luderer_energiewende_2025,agora_think_tanks_klimaneutrales_2024,thelen_paths_2025,boston_consulting_group_klimapfade_2021}.

Developing an efficient \ce{CO2} transport network is critical for large-scale CCS deployment.
This requires balancing transport capacity with CCS technology adoption.
Significant \ce{CO2} volumes are needed to justify pipeline investments, yet industrial emitters hesitate to adopt capture technologies without reliable transport infrastructure.
Given the lengthy planning, approval, and construction phases of infrastructure and industrial investments, coordinated development is essential \cite{kostka_large_2016}.

\noindent\textbf{Literature}

\noindent A growing body of literature examines individual steps of the carbon management chain.
On the storage side geological studies explore the storage potential in the North Sea, reservoir characteristics and feasible injection conditions to clarify where and at what rates \ce{CO2} can be sequestered \cite{gasanzade_subsurface_2025, rigby_storage_2023}.
On the capture side recent reviews synthesize the maturity and performance of major capture options across post-combustion, pre-combustion, oxy-fuel and emerging processes and their potential application in Germany \cite{wu_comprehensive_2024, borchers_comprehensive_2024}.
For transport and conditioning, the literature addresses compressor requirements, pressure regimes, material constraints, pipeline design, and broader feasibility questions, including public acceptance \cite{edelenbosch_reducing_2024,solomon_pipeline_2024,munkejord_co2_2016} and shipping requirements \cite{al_baroudi_review_2021}.
These studies provide the technical basis for source-to-sink optimization models that identify cost-efficient transport routes under uncertainty, including ship-based logistics and source-sink matching \cite{bjerketvedt_deploying_2022,bjerketvedt_optimal_2020,isoli_net_2025}.
At the same time, critical contributions argue that CCS should be prioritized for residual or hard-to-abate emissions and stress that large-scale deployment remains exposed to substantial technical, economic and project-realization risks \cite{malz_assessing_2025}.

Bottom-up and agent-based studies explore future carbon transport infrastructure, combining detailed datasets on industrial point sources, storage sites and exogenous transition pathways to derive explicit transport systems with a high level of detail \cite{biermann_role_2022,luo_simulation-based_2014,uddin_agent-based_2025}.
Recent work distinguishes between gaseous, dense-phase and liquefied \ce{CO2} transport and select among pipelines, ships, rail or road while also resolving transport conditions, compressor stations and concrete corridors \cite{anvari_carbon_2026,neuwirth_von_2025,european_commission_joint_research_centre_shaping_2024}.
While these studies provide a high level of detail, they are derived from exogenous scenario results  without feedback into the wider energy system.
In addition, they are framed as target-year designs for 2050 and therefore provide limited insight into start-up network decisions \cite{mengis_netzero_2022}.

Graph-theoretic methods provide a useful bridge between detailed infrastructure design and large-scale system optimization.
In energy research, graph theory has traditionally been applied primarily to electricity systems, where it is used to analyse network structure, dynamics, controllability, coherency and vulnerability \cite{ishizaki_graph-theoretic_2018}.
More recent work extends graph-based representations to energy routing and multi-layer energy networks, highlighting their usefulness for systems with coupled carriers and spatially distributed infrastructure \cite{guo_graph_2018,kazim_analysis_2025}.

Graph- and network-based methods have been used to design sparse energy transport systems, including pipeline layouts for distributed energy systems and multi-source heat networks \cite{cui_optimization_2021}.
In the context of carbon management, related optimization studies derive CO\textsubscript{2} transport and supply-chain configurations for Germany and compare alternative pipeline routing scenarios \cite{yeates_industrial_2024, leonzio_outlook_2019}.
Other work evaluates multimodal CO\textsubscript{2} transport options, including inland and offshore transport chains \cite{oeuvray_multi-criteria_2024}.

A key limitation of such corridor-focused studies is that infrastructure layouts are usually derived outside the wider energy system optimization.
Once a \ce{CO2} pipeline is built, additional emitters or removal options may connect to the same infrastructure over time, changing both its utilization rate and the value of early investments.
Yet integrated planning approaches remain limited.
Large-scale energy system studies often do not model explicit carbon transport infrastructure \cite{neumann_energy_2024,neumann_potential_2023, lindner_pypsa-_2025}.

Xiong et al. consider discrete \ce{CO2} pipelines projects that are awarded the status of projects of common and mutual interest by the EU alongside linear expansion of \ce{CO2} pipelines \cite{xiong_role_2025}.
Hofmann et al. explore a 2050 net-zero scenario where the European \ce{CO2} network is represented through continous linear investment variables and mostly serves to transport hard-to-abate emissions to sequestration sites \cite{hofmann_h2_2025}.
While this is computationally attractive, it tends to smooth lumpy infrastructure decisions across space and provides limited insight into specific pipeline segments.
Explicit mixed-integer formulations of routing and sizing, by contrast, become computationally demanding very quickly in large sector-coupled models \cite{li_mixed-integer_2022}.

The literature reveals a methodological gap between detailed bottom-up infrastructure planning and integrated energy system modelling.
Bottom-up studies can derive specific multimodal corridors and engineering designs, but they generally lack endogenous feedbacks from the evolving energy system and are typically focus on climate neutrality target years.
Integrated energy system models capture this feedback but usually provide only aggregate infrastructure requirements or continuous capacity expansions rather than discrete, interpretable topologies.

To address this gap, this work introduces a novel hybrid method that combines graph-based generation of candidate \ce{CO2} transport topologies with endogenous energy system optimization.
Rather than deriving a single deterministic network, multiple candidate topologies are generated and evaluated within an integrated model.
This allows us to consider discrete topologies for the first few pipelines that are built, while retaining the sector-coupled approach of the energy system model.
With this hybrid method we can address questions unanswerable in previous studies:

\begin{itemize}
    \item What should the first steps of building out Germany's \ce{CO2} network look like by 2035?
    
    \item How sensitive are energy system outcomes to different \ce{CO2} infrastructure topologies?

    \item How do early \ce{CO2} infrastructure decisions influence the transition towards climate neutrality?
\end{itemize}

\section{Methods}

The following section describes the graph theoretic approach to generate different network topologies (Sec.~\ref{sec:graph_theory}) before introducing the energy system model (Sec.~\ref{sec:esm}).
As well as a general description of PyPSA-DE, we focus on the endogenous implementation of industry processes and the \ce{CO2} streams within our model framework.

\subsection{Graph-based generation of candidate \ce{CO2} topologies}
\label{sec:graph_theory}

Candidate \ce{CO2} pipeline topologies are generated on a spatial graph whose nodes correspond to clustered model regions and whose edges correspond to potential pipeline corridors.
Each node $i \in V$ is assigned a capturable \ce{CO2} potential $p_i$.
This potential consists of industrial point sources and emissions from waste incineration and is only used to determine network topologies.
The actual \ce{CO2} flows and size of the pipelines are later determined by the energy system model.

A candidate graph

\begin{equation}
    G = (V,E)
\end{equation}

is constructed from the German regions and two adjacent international exits (the Netherlands and Denmark).
For every edge $e_{(i,j)}$ connecting two nodes $i$ and $j$, the physical edge length is denoted by $\ell_{ij}$.
If only the physical length $\ell_{ij}$ were used as the edge cost, the heuristic would prioritize compact networks composed of short connections, regardless of the \ce{CO2} point-source potential of the connected regions.
To favor corridors connecting high-potential \ce{CO2} regions first, each edge receives a weighted cost that is reduced for connections between sites with high \ce{CO2} capture potential

\begin{equation}
    c_{ij} = \frac{\ell_{ij}}
     {\left(1 + \tilde p_i + \tilde p_j\right)^\alpha},
\end{equation}

where

\begin{equation}
    \tilde p_i = \frac{p_i}{\max_{v \in V} p_v}
\end{equation}

is the normalized \ce{CO2} potential and $\alpha \in \mathbb{R}^{+}$ controls the importance of \ce{CO2} potential relative to distance.
As $\alpha \to 0$ the measure converges to a distance-minimizing network, while $\alpha > 0$ gives stronger preference to high-potential regions.

\noindent\textbf{Single-sink topology generation}

\begin{figure}[ht]
\centering
\begin{tikzpicture}[
    sink/.style={
        circle, draw=blue!60!black, fill=blue!18,
        minimum size=8mm, line width=1.2pt,
        font=\bfseries\small
    },
    node/.style={
        circle, draw=gray!60, fill=gray!10,
        minimum size=7mm, line width=0.8pt
    },
    source/.style={  % high-potential nodes
        circle, draw=orange!70!brown, fill=orange!15,
        minimum size=7mm, line width=0.8pt
    },
    chosen/.style={line width=1.8pt, black!80}, % -{Stealth[length=0pt]}
    unchosen/.style={gray!45, thin, dashed},
    label_style/.style={font=\scriptsize\itshape, text=gray!70}
]

% ── Nodes ────────────────────────────────────────────────────────────────────
\node[sink]   (s)  at (0,   0)   {$s$};
\node[source] (a)  at (1.8, 0)   {$p_0$};
\node[source] (b)  at (3.2, 1.1) {$p_1$};
\node[node]   (c)  at (3.2,-1.1) {$p_2$};
\node[source] (d)  at (4.8, 1.3) {$p_3$};
\node[node]   (e)  at (4.9,-0.9) {$p_4$};

% ── Unchosen edges (draw first, below chosen) ────────────────────────────────
\draw[unchosen] (b)--(c);
\draw[unchosen] (b)--(e);
\draw[unchosen] (c)--(d);
\draw[unchosen] (d)--(e);
\draw[unchosen] (c)--(e);
\draw[unchosen] (a)--(c);

% ── Chosen tree ──────────────────────────────────────────────────────────────
\draw[chosen] (s)--(a);
\draw[chosen] (a)--(b);
\draw[chosen] (b)--(d);

% ── Edge weight labels on chosen edges ───────────────────────────────────────
\node[label_style] at (0.9,  0.18) {$c_{s0}$};
\node[label_style] at (2.4,  0.75) {$c_{01}$};
\node[label_style] at (3.95, 1.45) {$c_{13}$};

% ── Ellipsis gap ─────────────────────────────────────────────────────────────
\node at (5.8, 1.3) {\Large$\cdots$};
\node at (6, -0.8) {\Large$\cdots$};

% ── Isolated nodes i and j with candidate edge ───────────────────────────────
\node[node] (ni) at (6.6,  1.3) {$p_i$};
\node[node]   (nj) at (8.0, 0.3) {$p_j$};

\draw[unchosen] (ni)--(nj);
\node[label_style] at (7.2, 0.5) {$c_{ij}$};

\end{tikzpicture}
\caption{Single-sink topology construction: Starting from the sink node~$s$ (blue), the algorithm greedily selects edges (solid) to connect nodes with a given \ce{CO2} point source potential~$p_i$ (orange) within the length limit.
Grey dashed edges are candidate connections not chosen in this topology.}
\label{tikz:single_sink}
\end{figure}
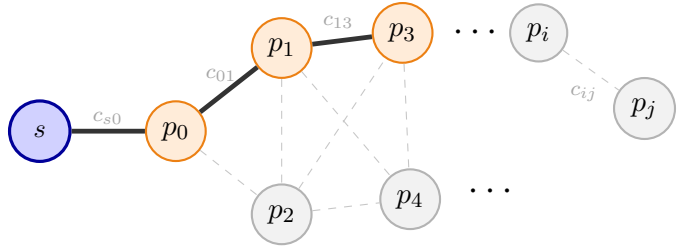

In the single-sink cases, the topology is generated with a Prim-style greedy tree-growth heuristic on the weighted candidate graph.
Prim's algorithm constructs a minimum spanning tree by iteratively adding the least-cost edge that connects the already connected part of the graph to an unconnected node \cite{prim_shortest_1957}.
Here, this principle is adapted to a budget-constrained network design problem.
Beginning at that sink node, the algorithm scans edges in increasing order of weighted cost $c_{ij}$ and adds an edge if it connects exactly one already connected node to one unconnected node (Fig~\ref{tikz:single_sink}).
Edges that would connect two already connected nodes are skipped to avoid cycles, while edges connecting two unconnected nodes are ignored because they do not extend the current sink-rooted component.
This grows a tree outward from the selected sink until a total length budget
$L^{\max}$ is exhausted.
The resulting topology is not necessarily a minimum spanning tree of the full graph, but a rooted, budget-constrained approximation that prioritizes short corridors and high-potential \ce{CO2} source regions.
For interconnectors only half of the length is considered towards this budget.

\begin{equation}
    \sum_{(i,j)\in T} \ell_{ij}^{\mathrm{budget}} \leq L^{\max}.
\end{equation}

\noindent\textbf{Multi-sink topology generation and perturbed topologies}

The multi-sink case extends this heuristic to several possible sink regions.
The start set contains all available sink nodes,

\begin{equation}
    S = \{s_{0}, \dots, s_{n}\}.
\end{equation}

Depending on the scenario, the algorithm initializes one or more independent sink-rooted strands.
If only one strand is initialized, the procedure is equivalent to the single-root case, except that additional sink nodes may be connected later.
If several strands are initialized, multiple sink-rooted components are allowed to grow in parallel.

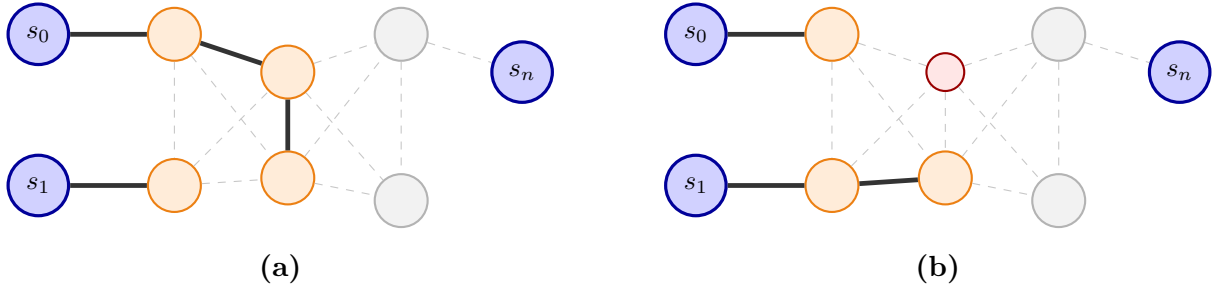
\begin{figure}[h]
\centering
\tikzset{
    sink/.style={
        circle, draw=blue!60!black, fill=blue!18,
        minimum size=8mm, line width=1.2pt,
        font=\bfseries\small
    },
    node connected/.style={
        circle, draw=orange!70!brown, fill=orange!15,
        minimum size=7mm, line width=0.8pt
    },
    node unconnected/.style={
        circle, draw=gray!60, fill=gray!10,
        minimum size=7mm, line width=0.8pt
    },
    node reduced/.style={
        circle, draw=red!60!black, fill=red!10,
        minimum size=5mm, line width=0.8pt
    },
    chosen/.style={line width=1.8pt, black!80},
    unchosen/.style={gray!45, thin, dashed},
    label_style/.style={font=\scriptsize\itshape, text=gray!70}
}

% ─────────────────────────────────────────────────────────────────────────────
% (a) Base topology
% ─────────────────────────────────────────────────────────────────────────────
\begin{tikzpicture}
    \node[sink]             (s1) at (0,    1.0) {$s_{0}$};
    \node[sink]             (s2) at (0,   -1.0) {$s_{1}$};
    \node[sink]             (s3) at (6.4,  0.5) {$s_{n}$};
    \node[node connected]   (a)  at (1.8,  1.0) {};
    \node[node connected]   (b)  at (1.8, -1.0) {};
    \node[node connected]   (c)  at (3.3,  0.5) {};
    \node[node connected]   (d)  at (3.3, -0.9) {};
    \node[node unconnected] (e)  at (4.8,  1.0) {};
    \node[node unconnected] (f)  at (4.8, -1.2) {};

    \draw[unchosen] (e)--(s3);
    \draw[unchosen] (a)--(b);
    \draw[unchosen] (a)--(d);
    \draw[unchosen] (b)--(c);
    \draw[unchosen] (b)--(d);
    \draw[unchosen] (c)--(e);
    \draw[unchosen] (e)--(f);
    \draw[unchosen] (d)--(f);
    \draw[unchosen] (c)--(f);
    \draw[unchosen] (d)--(e);

    \draw[chosen] (s1)--(a);
    \draw[chosen] (s2)--(b);
    \draw[chosen] (a)--(c);
    \draw[chosen] (c)--(d);

    % panel label
    \node at (3.2, -2.1) {\textbf{(a)}};
\end{tikzpicture}
\hspace{1.2cm}
%
% ─────────────────────────────────────────────────────────────────────────────
% (b) Reduced potential at c
% ─────────────────────────────────────────────────────────────────────────────
\begin{tikzpicture}
    \node[sink]             (s1) at (0,    1.0) {$s_{0}$};
    \node[sink]             (s2) at (0,   -1.0) {$s_{1}$};
    \node[sink]             (s3) at (6.4,  0.5) {$s_{n}$};
    \node[node connected]   (a)  at (1.8,  1.0) {};
    \node[node connected]   (b)  at (1.8, -1.0) {};
    \node[node reduced]     (c)  at (3.3,  0.5) {};   % reduced potential
    \node[node connected]   (d)  at (3.3, -0.9) {};
    \node[node unconnected] (e)  at (4.8,  1.0) {};
    \node[node unconnected] (f)  at (4.8, -1.2) {};

    \draw[unchosen] (e)--(s3);
    \draw[unchosen] (a)--(b);
    \draw[unchosen] (a)--(d);
    \draw[unchosen] (b)--(c);
    \draw[unchosen] (a)--(c);   % a-c now unchosen
    \draw[unchosen] (c)--(d);   % c-d now unchosen
    \draw[unchosen] (c)--(e);
    \draw[unchosen] (e)--(f);
    \draw[unchosen] (d)--(f);
    \draw[unchosen] (c)--(f);
    \draw[unchosen] (d)--(e);

    \draw[chosen] (s1)--(a);
    \draw[chosen] (s2)--(b);
    \draw[chosen] (b)--(d);     % rerouted through d instead

    % panel label
    \node at (3.2, -2.1) {\textbf{(b)}};
\end{tikzpicture}

\caption{Multi-sink topology construction with source scenario variation.
\textbf{(a)}: the network grows from sink nodes $s_0, s_1, \ldots,
s_n$ (blue); orange nodes are connected within the pipeline budget (solid
edges) and grey nodes remain unconnected (dashed edges are unchosen
candidates).
\textbf{(b)}: perturbed \ce{CO2} potential at on enode (red):
the edge connecting a node becomes less attractive which leads to an alternative route}
\label{tikz:multi-sink}
\end{figure}

Before expanding towards further source regions, the algorithm first attempts to connect additional sink nodes to the already initialized components.
For this step, shortest weighted paths are computed from the current set of connected nodes to the remaining sink nodes using a multi-source variant of Dijkstra's algorithm \cite{dijkstra_note_1959}.
Among the feasible paths, the path with the lowest weighted cost is selected if its additional effective length does not exceed the remaining length budget.
After this sink-connection step, the algorithm again applies the Prim-style greedy expansion described above.
The resulting topology can therefore be a single rooted tree or, when several independent strands are allowed, a forest of sink-oriented branches (see Fig.~\ref{tikz:multi-sink}~(a)).

Additional topology variants are generated by perturbing the \ce{CO2} point-source potentials of already connected regions.
These perturbations reflect uncertainties about how much \ce{CO2} will be available to capture in a specific region based on the decision of individual industrial plants.
The \ce{CO2} point-source potential of up to three connected nodes is reduced before the edge weights are recalculated.
This changes the attractiveness of corridors passing through these regions and can force the heuristic to select alternative routes.

\subsection{Energy system model}
\label{sec:esm}

We use the sector-coupled energy system model PyPSA-DE, which represents energy and feedstock demand across the electricity, heating, transport, industry and agricultural sectors.
The model includes a broad range of technologies for the generation, conversion and storage of energy, and optimizes both capacity expansion and operation under the objective of minimizing total annual system costs while meeting demand and complying with emission constraints.

The model places special focus on Germany while including its electricity neighbors as well as Spain and Italy, which are promising locations for future hydrogen electrolysis due to high renewables availability.

Additional details on the model formulation are provided in Appendix~A and can be found in literature \cite{lindner_kopernikus-projekt_2025,lindner_pypsa-_2025}.

\noindent\textbf{Industry representation}

\noindent With hard-to abate emissions especially occuring in the industrial sector, the model explicitly represents major industrial production processes and their associated \ce{CO2} emissions.
These include the production of methanol, ammonia, steel and cement.

Methanol production can occur via hydrogen- and biomass-based pathways, while ammonia production is represented through a Haber-Bosch process using hydrogen and electricity for an air-separation unit to provide nitrogen.
Primary steel production can either occur via a coal-based blast furnace route or via direct reduction using hydrogen or natural gas in combination with an electric arc furnace.
Cement production consists of clinker production from limestone and a subsequent finishing process mixing clinker with gypsum.
Thermal energy demand for cement kilns can be supplied via biomass, waste and natural gas.
The model explicitly represents carbon capture options for energy-intensive industries.
The model can choose whether to equip coal-based blast furnace steel plants and natural gas-based direct reduction plants with post-combustion carbon capture.
Likewise, \ce{CO2} capture can be applied to cement kiln processes.
The model does not include oxyfuel carbon capture processes.
An overview of the endogenously modelled industry processes is given in Appendix~A.

Industrial production levels are linked to exogenously specified demand trajectories.
Demand assumptions follow reduced industrial demand projections for the year 2035 consistent with the ``Further measures'' scenario published by the \textit{German Environment Agency}.
According to these projections, annual industrial final energy demand declines from 614~TWh in 2025 to 576~TWh in 2035 \cite{forster_treibhausgas-projektionen_2025}.

A large share of industrial subsectors is not represented by explicit process models.
These include high-value chemicals, pharmaceuticals, ceramics, glass, pulp and paper, food production, aluminium production and other manufacturing industries.
For these sectors, energy and feedstock demands are represented in an aggregated form at regional level based on secondary energy consumption, including electricity, coal, biomass, gas, hydrogen, low-temperature heat, oil and methanol.
Associated process and feedstock-related emissions are also treated as aggregated regional emission point sources.

The existing industrial fleet for steel, cement, methanol and ammonia production with their building year is represented explicitly using plant-level datasets \cite{global_energy_monitor_global_2025,global_energy_monitor_global_2025-1,neuwirth_modelling_2024}.
For steel plants, relining and retirement dates are included where available.
Where no retirement date is available, plants are assumed to remain in operation until the end of their technical lifetime.
This enables a brownfield representation of industrial transition pathways and allows the model to capture the gradual replacement of existing industrial assets by low-carbon alternatives.
A validation of the industrial capacities is given in Appendix~A.

\noindent\textbf{Carbon Management}

\begin{figure}
    \centering
    \includegraphics[width=1.0\linewidth]{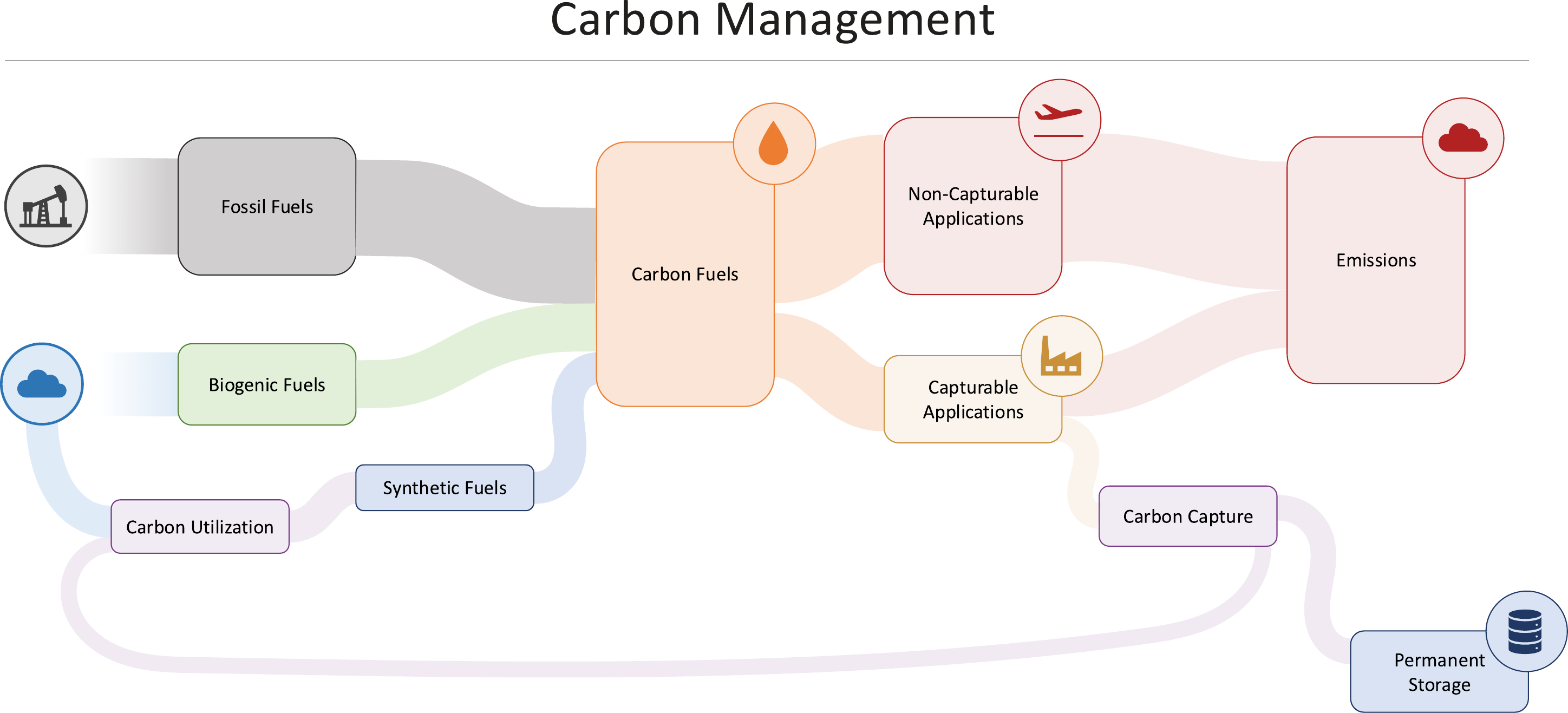}
    \caption{CO$_2$ streams represented in the model.
    Carbon enters the system with a negative footprint if it originates from biogenic sources.
    Once a carbon-based fuel is used in a non-capturable application, the associated CO$_2$ is emitted to the atmosphere.
    For capturable point sources, the model can either emit or capture the resulting CO$_2$.
    Captured CO$_2$ can subsequently be utilized for synthetic fuel production or permanently stored.
    The figure is a stylized representation of the carbon accounting and does not include process losses or leakage during capture or utilization.
    }
    \label{fig:carbon_management}
\end{figure}

\noindent The representation of carbon flows in the model is illustrated in Fig.~\ref{fig:carbon_management}.
The model distinguishes between capturable point-sources and distributed sources that are assumed not to be economically capturable.

Non-capturable emissions include emissions from gas and oil boilers to meet heating demand as well as emissions from the transport sector.
These emissions can only be avoided by using synthetic fuel produced from hydrogen and renewable \ce{CO2} or by electrification with the latter being an exogenous choice based on current trends.
Capturable point-source emissions include emissions from gas-, biomass- and waste-fired power or CHP plants as well as industrial process emissions.
Industrial capturable sources comprise cement production, steelmaking routes and emissions associated with the industrial use of gas and biomass.

Captured \ce{CO2} can either be permanently stored or utilized by synthesizing methanol, Fischer-Tropsch or methane.
The allocation between utilization and sequestration is determined endogenously within the model.
Biogenic \ce{CO2} emissions are treated as carbon-neutral.
Sequestration therefore is counted as negative, which can offset residual fossil emissions elsewhere in the system.
Direct air capture (DAC) is not available in the 2035 model horizon.
Although DAC is expected to become an important carbon removal technology in the long term, it is not expected to be deployed at scale in Europe by 2035.

The model is constrained by national carbon budgets consistent with European and German climate targets.
Non-\ce{CO2} emissions are subtracted from the available carbon budget, while land-use, land-use change and forestry are not represented explicitly.
Emissions from international bunkers are excluded from the German carbon budget.
For sequestration exhausted oil and gas fields as well as aquifer storage are available in the North Sea.
We assume a total volume of \SI{100}{Mt/a} that can be permanently stored in 2035.
Techno-economic parameters for carbon management technologies are listed in Appendix~B.

\noindent\textbf{\ce{CO2} infrastructure}

\noindent \ce{CO2} transport is represented explicitly in order to capture the infrastructure requirements associated with carbon capture and sequestration.
Captured \ce{CO2} is transported in dense phase via pipelines only.
We take into account pipeline and compressor investment costs and the electricity demand required for \ce{CO2} compression to dense-phase.

Given the limited spatial resolution of the model, \ce{CO2} transport is represented by inter-regional pipeline corridors between clustered model regions rather than by detailed local pipeline networks.
Pipeline distances are approximated using the centroid distance between regions multiplied by a routing factor of 1.25 to account for deviations from the shortest-path connection.

To represent infrastructure routing decisions, candidate \ce{CO2} transport corridors are generated beforehand using the graph-based heuristic described in Section~\ref{sec:graph_theory}.
Only these predefined pipelines segments are included in the energy system model and are optimized continously.
The model can choose to capture, transport and utilize or permanently store \ce{CO2} along these segments.
Large-scale energy system models can give insights into overall infrastructure requirements but fail to produce shortest path topologies.
The graph-theoretic approach therefore determines the candidate network topology, while the energy system model determines its utilization and optimal capacity expansion.
Instead of prescribing a fixed pipeline diameter for all selected corridors, pipeline capacities remain endogenous.
This allows the model to distinguish between heavily and lightly utilized transport corridors and provides insight into which pipeline segments are chosen within the model framework.
Electricity and hydrogen transmission capacities, as well as \ce{CO2} transport infrastructure outside Germany remain continuously expandable.

Geological \ce{CO2} sequestration is represented using spatially explicit storage potentials.
In general, sequestration is assumed to occur via pipeline transport to storage sites.
An exception is made for the German North Sea, where offshore storage is not assumed to be commercially available by 2035.
This is in line with the current projects of common and mutual interest of the European Union even though research is conducted to explore the storage volume in the German North Sea \cite{european_commission_commission_2025, geostor-konsortium_co-speicherung_2025}.
For these North Sea regions, sequestration is therefore only possible via shipping to the \textit{Northern Lights} project in Norway.

\noindent\textbf{Model setup}

\noindent A total of 89 nodes are distributed across the model scope with increased spatial resolution in Germany and in coastal regions with access to \ce{CO2} storage sites (see Appendix~A).
We use a temporally segmented representation with an average temporal resolution of four hours.
This allows a higher temporal resolution during periods that are particularly challenging for the energy system while lower temporal resolution is used during less critical periods in order to reduce computational complexity.
The chosen spatial and temporal resolution allows us to investigate system behavior at a high level of detail while managing computational time.

The model is solved as a brownfield optimization for the years 2025 and 2035.
Existing electricity generation, heating technologies, industrial production capacities and infrastructure assets are included as the starting point of the optimization.
To be able to compare our results, we additionally optimize a scenario without any \ce{CO2} pipeline infrastructure across the entire model scope.
In that case, CCS is only available to regions closest to the offshore sequestration site.

To analyse the long-term implications of infrastructure decisions in 2035, the topologies are subsequently optimized for the year 2050 under climate neutrality constraints.
\ce{CO2} infrastructure can be optimized freely to investigate whether infrastructure choices in 2035 create lock-in effects that increase long-term system costs or constrain later decarbonisation pathways.

\section{Results}

\subsection{Network topologies}

We apply the graph-theoretic topology-generation method to construct a set of candidate \ce{CO2} transport networks for Germany.
The topology ensemble varies four main parameters:
\begin{itemize}
    \item the sink set which includes shipping from Germany to the \textit{Northern Lights} project in Norway and connections to Denmark and the Netherlands: $S=\{s_{\mathrm{DE}},s_{\mathrm{DK}},s_{\mathrm{NL}}\}$
    \item the source-weighting parameter which is varied between $\alpha=0.5$ and $\alpha=2$
    \item and the maximum network length which is set to $L^{\max}\in\{\SI{500}{km},\SI{1000}{km},\SI{1500}{km}\}$.
\end{itemize}

The network-length budget should be interpreted as a stylized measure of the overall extent of the transport network rather than the physical pipeline length.
Pipeline routes are represented by straight-line connections between regional centroids, whereas real-world pipelines are routed along existing infrastructure corridors and avoid densely populated areas and protected areas.
Consequently, the topology-generation procedure is intended to compare relative infrastructure extents rather than predict exact pipeline routing.
For each multi-sink configuration, up to two perturbation steps representing demand uncertainty are applied to generate alternative routing options while limiting the total number of topology variants.
Duplicate topologies are removed before determining the final set of topologies.

The individual topologies are given in Appendix~C.
Fig.~\ref{fig:topologies_compressed} summarizes the resulting topologies by showing the selection frequency of each pipeline segment across all generated topologies for a given length limit.
A selection frequency of 100~\% indicates that a pipeline segment is present in every generated topology of the corresponding ensemble, whereas low frequencies indicate corridors that only appear in a small subset of topologies.

For maximum network lengths of \SI{500}{km} and \SI{1000}{km}, the resulting topologies exhibit strong variability and only few pipeline segments achieve high selection frequencies. 
Under restrictive infrastructure budgets the topology depends mostly on the starting sink node(s).
In these constrained cases, the network length is often insufficient to simultaneously connect major industrial source regions.
As the maximum network length increases to \SI{1500}{km}, several recurring corridors emerge with high selection frequencies.

Across all length limits, the generated topologies concentrate in north-western Germany.
This results both from the location of the considered sink regions in northern and western Germany and from the high concentration of industrial \ce{CO2} sources in these regions.

\begin{figure}[htbp]
    \centering
    \includegraphics[width=0.8\textwidth]{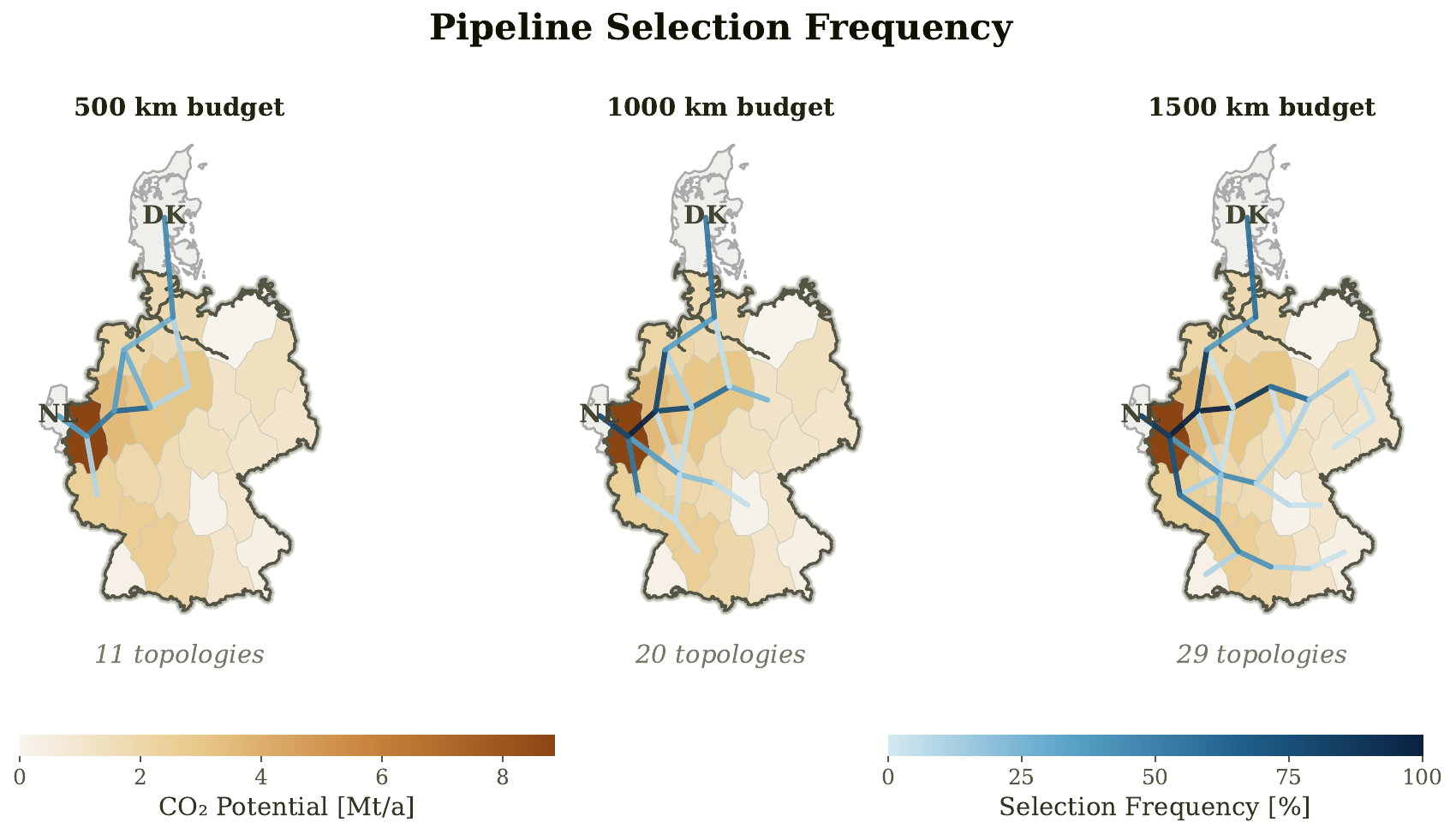}
    \caption{
    Selection frequency of pipeline segements. High frequencies (dark blue) translate to corridors that repeatedly emerge across different sink assumptions, weighting parameters and perturbation.
    The color of each region represents the point source potential for \ce{CO2} volumes from industrial sources and waste incineration facilities.
}
\label{fig:topologies_compressed}
\end{figure}

\subsection{Energy system behavior}

The following section evaluates the impact of alternative \ce{CO2} transport topologies within an integrated European energy system model.
Because investment and dispatch decisions are optimized jointly across Europe, total system costs cannot be uniquely assigned to individual countries.
Cross-border infrastructure investments and energy flows may compensate for missing German investments while still meeting demand.

Instead of discussing total system costs, we analyse the cost of meeting the German demand which include electricity, hydrogen, heat, biomass, liquid hydrocarbons, methanol, HVC, ammonia, steel and cement.
From our optimization results, we extract the Langragian multipliers $\lambda_{i,t}$, which represent the marginal price for an energy carrier or feedstock in region $i$ and time step $t$.
When multiplied by the corresponding consumption $d$, we obtain the consumer costs $c_{cons}$

\begin{equation}
    c_{cons}  = \sum_{i \in I_{DE}} \sum_{t \in T} \lambda_{i,t} \cdot d_{i,t}
\end{equation}

\noindent\textbf{Economic Value of Carbon Infrastructure in 2035}

\begin{figure}[htbp]
    \centering
    \includegraphics[width=0.5\textwidth]{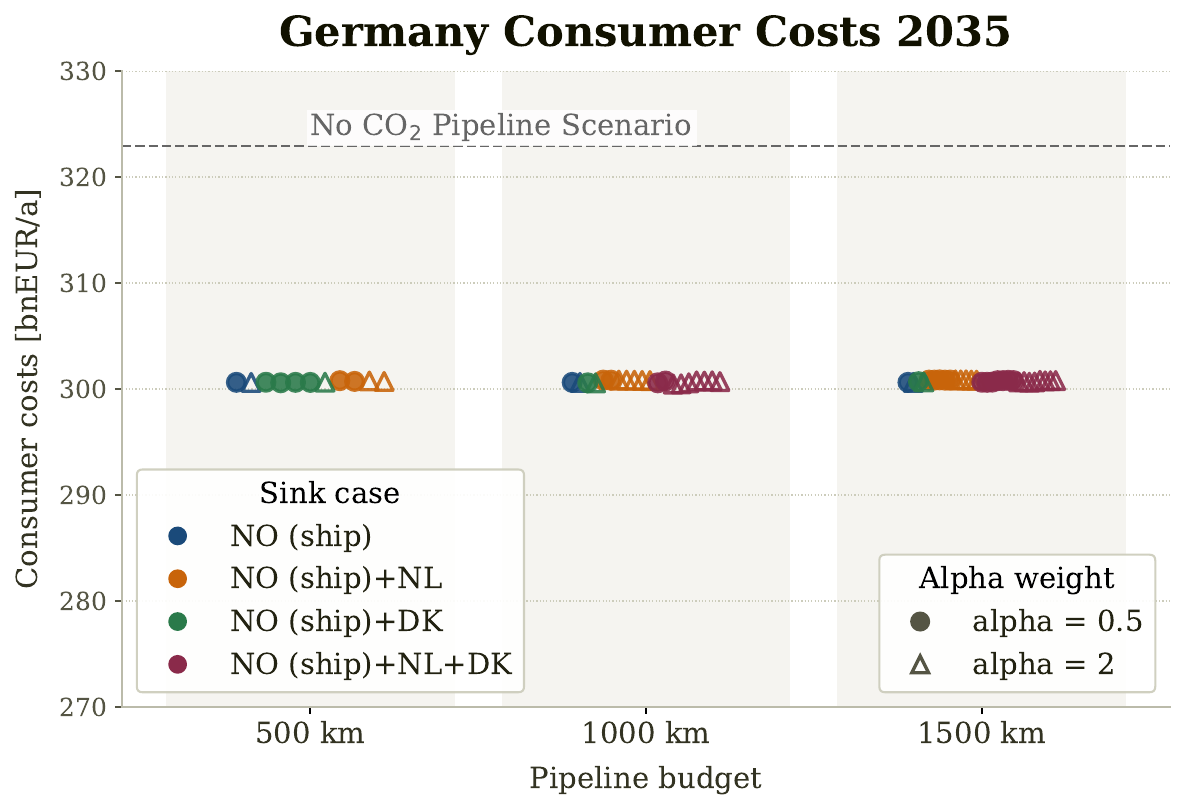}
    \caption{German consumer costs for different carbon infrastructure topologies in 2035. Each symbol represents one topology run and is grouped for different pipeline length limits. The color indicates the available sinks and the symbol the $\alpha$ weighting of \ce{CO2} point sources for the development of the topology.}
    \label{fig:cc_2035}
\end{figure}

\noindent Figure~\ref{fig:cc_2035} shows the resulting German consumer costs for all topology variants.
Across all scenarios, consumer costs average \SI{300.7}{bnEUR/a}, with only small differences of $\pm 0.2$ \SI{}{bnEUR/a} between network topologies.
The maximum pipeline length and the selected sink configuration have only a limited influence on consumer costs in 2035.

By comparison, a scenario without any \ce{CO2} transport infrastructure in 2035 and permanent sequestration only being available to regions closest to the offshore storage sites results in German consumer costs of \SI{322.9}{bnEUR/a}.
This corresponds to savings of \SIrange{22.0}{22.4}{bnEUR/a} (\SIrange{6.8}{6.9}{\percent}) once a \ce{CO2} transport network is available.
The higher consumer costs are driven by an increase in the shadow price of the European carbon constraint from approximately \SI{392}{EUR/tCO2} to \SI{482}{EUR/tCO2}.
This indicates that, without access to \ce{CO2} transport infrastructure, the energy system loses one of its lowest-cost mitigation options and must instead rely on more expensive decarbonization pathways to satisfy the same emissions target.
A more detailed breakdown of the consumer cost savings is provided in Appendix~D and a detailed comparison between a system with and without \ce{CO2} pipelines is discussed in Appendix~E.
The limited sensitivity to network topology suggests that the availability of \ce{CO2} transport infrastructure is more important than its precise layout during the early deployment phase.

\noindent\textbf{German Carbon Balance}

\begin{figure}[htbp]
    \centering
    \includegraphics[width=1.0\textwidth]{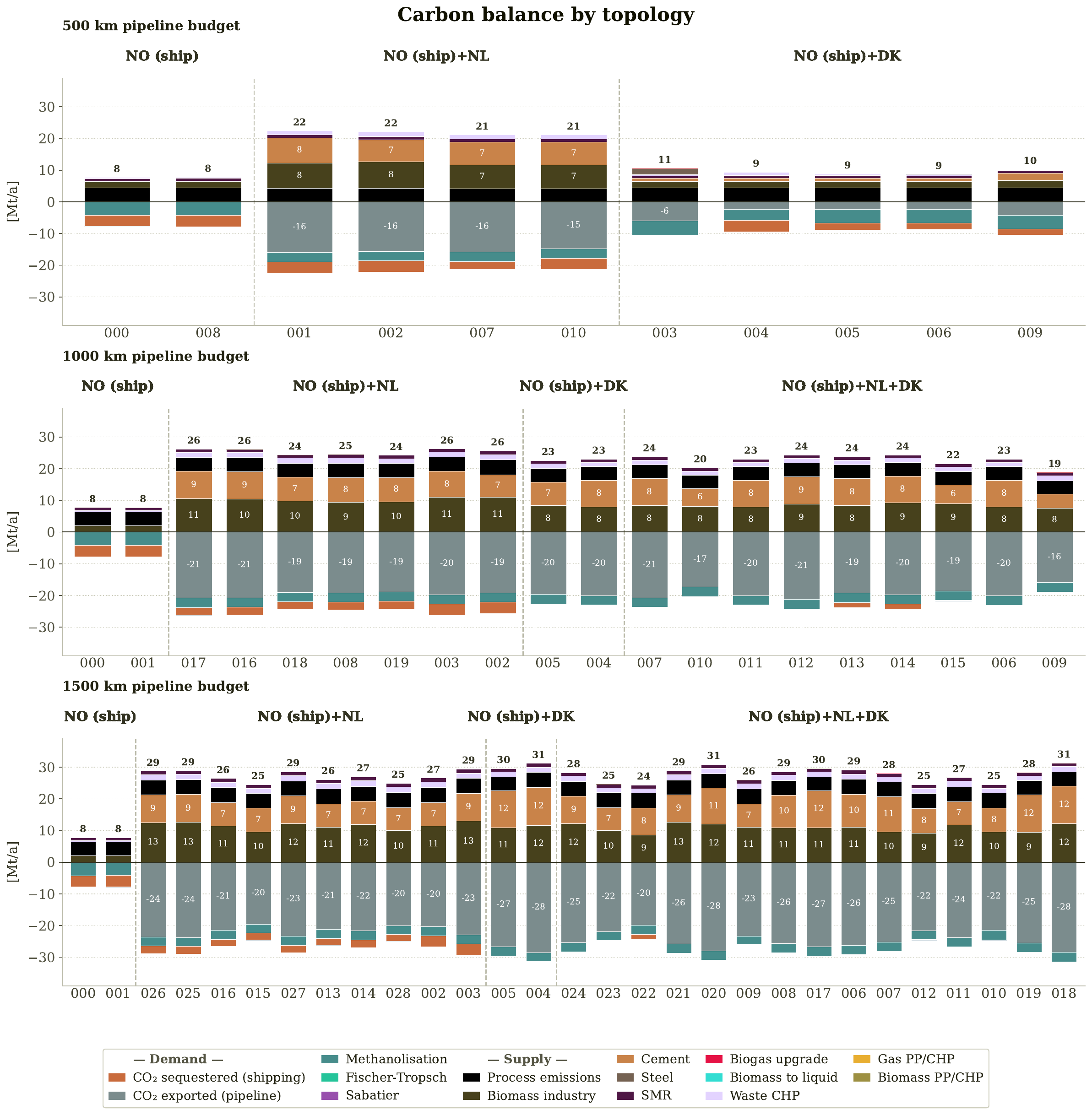}
    \caption{Carbon balances for different network lengths and sink cases in 2035. The different topologies are numbered along the x-axis. Positive values are showing the source of captured \ce{CO2} while negative values show how the \ce{CO2} stream is utilized or permanently stored.}
    \label{fig:carbon_balance}
\end{figure}

\noindent Fig.~\ref{fig:carbon_balance} shows the German carbon balance for each of the topologies.
With a pipeline length budget of \SI{500}{km}, the selected sink location strongly influences the amount of \ce{CO2} transported for sequestration.
If sequestration is only available via shipping, approximately \SI{8}{Mt/a} of \ce{CO2} are captured with only part of this volume being permanently stored due to the comparatively high cost of sequestration.
Carbon utilization through methanol synthesis becomes more attractive than in other cases.
Providing an interconnector to the Netherlands substantially increases sequestration volumes to \SIrange{21}{22}{Mt/a}.
This is primarily due to the proximity of the Dutch sink route to the high capturable \ce{CO2} potentials in North Rhine-Westphalia.
By contrast, topologies with access to Denmark reach only about \SIrange{11}{12}{Mt/a}, reflecting the lower availability of suitable point sources close to the Danish interconnector under the restricted length budget.

For larger length budgets, captured \ce{CO2} volumes remain relatively constant as long as Germany is connected to Denmark or the Netherlands.
If the only option is transport by ship for permanent storage, the high cost of sequestration drives the system to use alternative mitigation options.
Across the \SI{1000}{km} and \SI{1500}{km} topology budgets, total captured \ce{CO2} volumes reach up to \SI{31}{Mt/a} compared to the \SI{100}{Mt/a} European sequestration limit.
The most attractive carbon capture sources are industrial point sources, including process emissions, cement production and biomass-based carbon dioxide removal.
These are complemented by smaller capture volumes from waste-fired backup capacities.
Due to the low flue gas concentration of \ce{CO2} in gas-fired power and CHP plants, which lead to higher investment costs per ton of \ce{CO2} captured, we do not observe the deployment of carbon capture units in combination with gas-fired plants.
However, small share of \ce{CO2} are contributed by blue hydrogen.

Emissions in regions without access to \ce{CO2} transport can be offset by carbon dioxide removal from biomass-based processes in connected regions.
As a result, \ce{CO2} infrastructure provides a measurable system benefit, but differences between individual topology realizations remain comparatively small once major source regions and sink routes are connected.

Increasing the maximum network length generally increases the amount of captured \ce{CO2}.
This effect is partly driven by the continuous representation of pipeline expansion within the energy system model.
Once a pipeline corridor is available, the optimization can construct very small pipeline capacities to connect remote emission sources if their capture costs are sufficiently attractive. In practice, such small pipeline capacities may not be technically or economically feasible due to minimum pipeline diameters and economies of scale.
Consequently, longer topology budgets enable the model to connect an increasing number of dispersed capture sources, particularly process emissions, cement production and biomass-based carbon dioxide removal.
The resulting transport network resembles a collection system in which small upstream branches aggregate \ce{CO2} volumes that are transported through progressively larger corridors towards the available sink regions.

\noindent\textbf{Utilization of Pipeline segments}

\begin{figure}[htbp]
    \centering
    \includegraphics[width=0.8\textwidth]{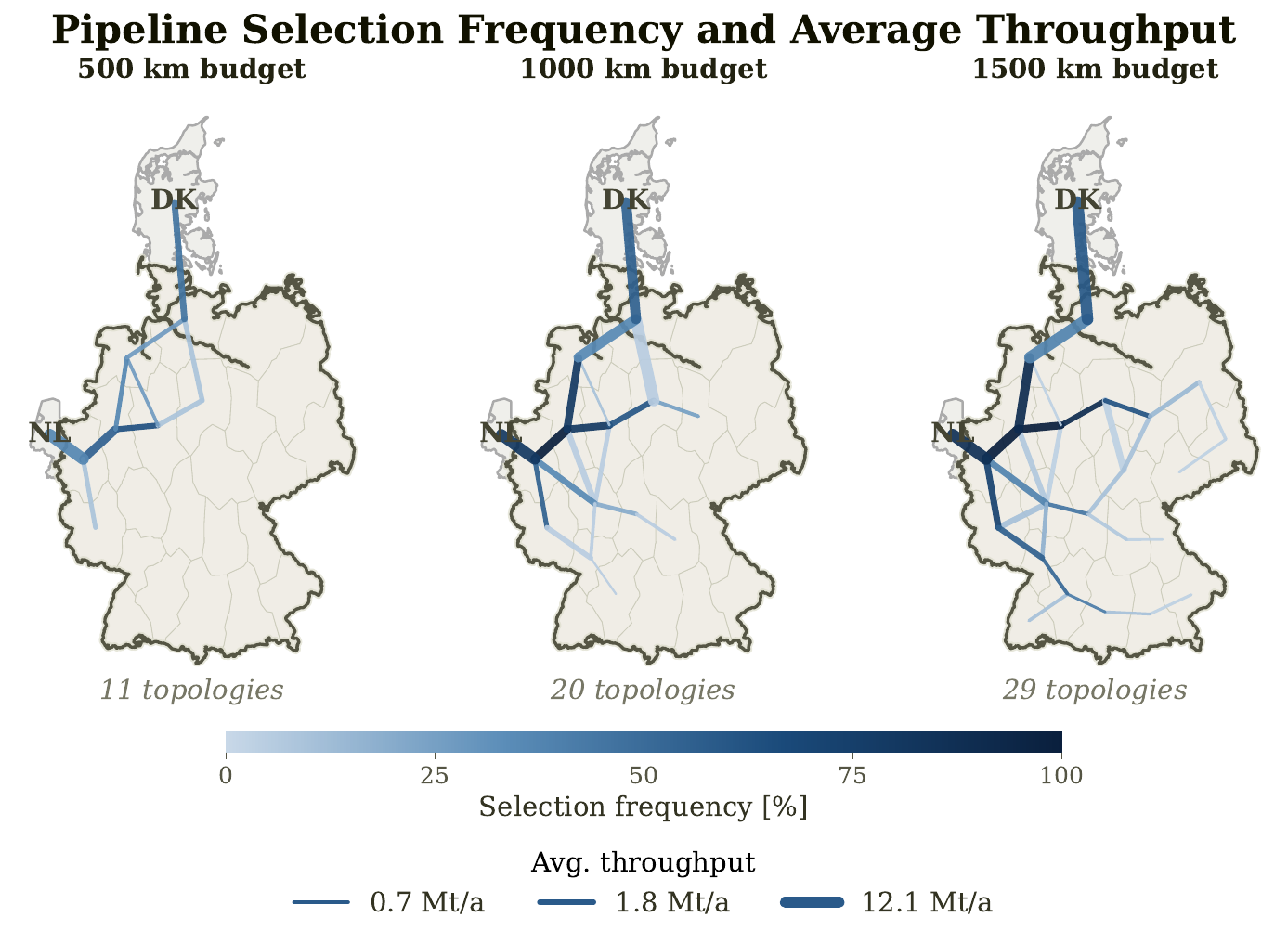}
    \caption{Selction frequency of pipeline segments from graph theory (color) and if chosen average throughput of CO2 (thickness) in 2035.}
    \label{fig:topologies_throughput}
\end{figure}

\noindent Fig.~\ref{fig:topologies_throughput} combines the selection frequency of pipeline segments with their operational relevance by including the average transported \ce{CO2} volume.
Across all network length limits, the highest throughput corridors are concentrated in north-western Germany.
These corridors connect the major industrial point sources in North Rhine-Westphalia to the available sink regions in the Netherlands and Denmark.
For the shortest network length of \SI{500}{km}, the connection towards the Netherlands emerges as the dominant transport corridor, indicating that access to Dutch transport and storage infrastructure provides a particularly valuable outlet for captured \ce{CO2} under constrained infrastructure budgets.

As the maximum network length increases to \SI{1000}{km} and \SI{1500}{km}, additional corridors emerge linking the Danish and Dutch sink regions through northern Germany and even reach Saxony-Anhalt and Hesse.
Despite the larger available network budget, the resulting transport network does not expand uniformly across Germany.
With increasing network length budget, more peripheral pipeline segments reaching to South and East Germany are included but the energy system model does not choose to utilize these.

\noindent\textbf{Long-term effects}

\noindent To investigate the long-term implications of early infrastructure decisions, the analysis is extended to the year 2050, when Germany and all neighbouring countries have reached climate neutrality.
The \ce{CO2} pipeline topologies selected in 2035 are retained, while additional pipeline expansion is allowed without a predefined length constraint.
In contrast to 2035, all geological sequestration sites are assumed to be available, including offshore storage in the German North Sea.
The larger European scope also enables Germany to act as a transit country for \ce{CO2} flows from neighbouring landlocked countries such as the Czech Republic, Austria and Switzerland.

\begin{figure}[htbp]
    \centering
    \includegraphics[width=1.0\textwidth]{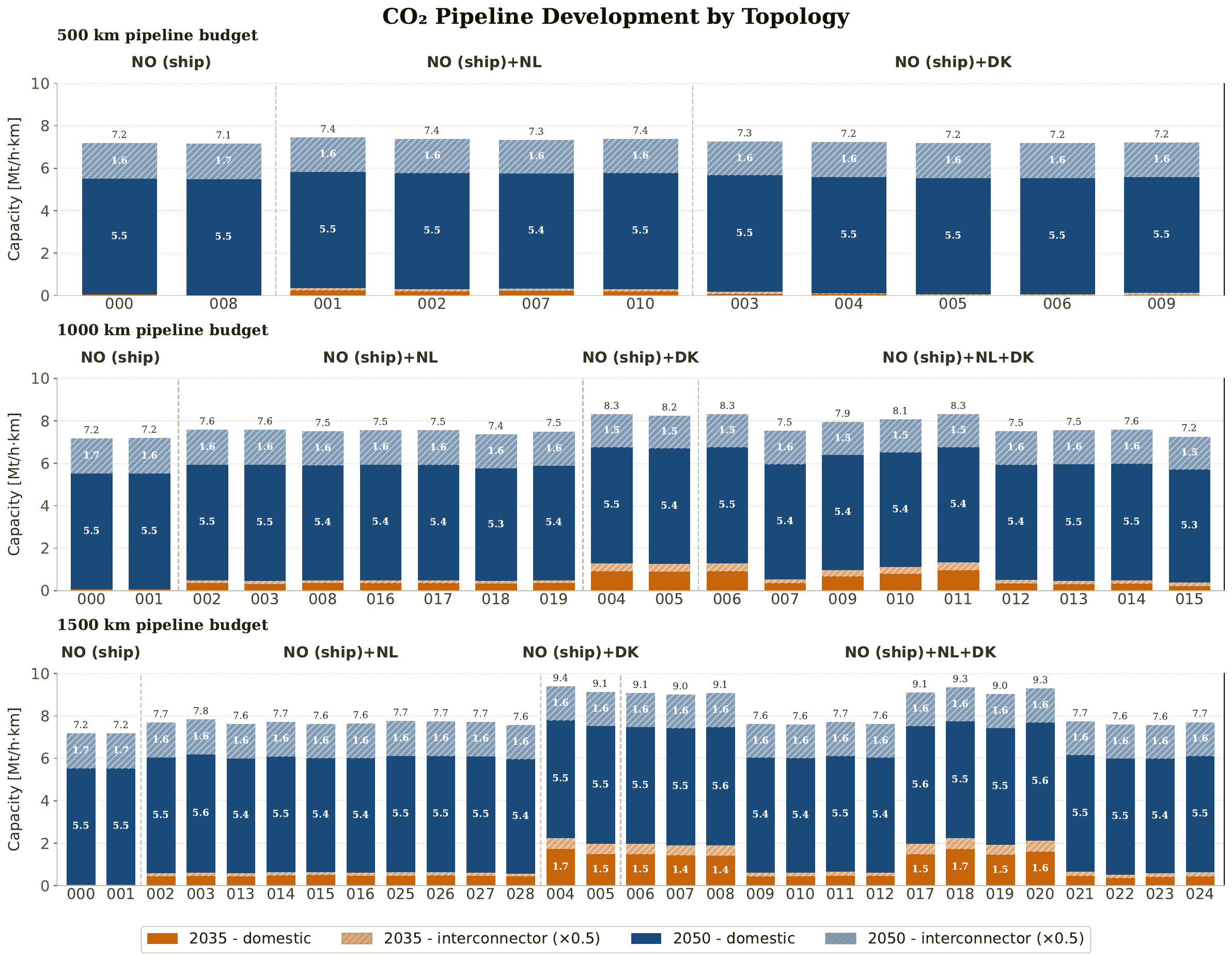}
    \caption{Carbon infrastructure build out until 2035 (orange) and between 2035 and 2050 (blue) across different topologies in 2035.}
    \label{fig:expansion}
\end{figure}

Figure~\ref{fig:expansion} shows the \ce{CO2} pipeline infrastructure deployed by 2035 and the additional expansion required between 2035 and 2050.
Infrastructure requirements are quantified as the sum of installed capacity multiplied by segment length across all pipeline segments.
Under the tightest 2035 network budget, pipeline deployment remains comparatively limited, particularly in the shipping-only and Danish sink configurations.
This is consistent with the low German capture volume of approximately \SI{10}{Mt/a} in the Norway and Denmark cases in 2035.
Despite these differences in the initial build-out, total infrastructure requirements converge by 2050, reaching approximately \SIrange{7.1}{7.4}{Mt/h\cdot km} for the most constrained early networks.

For larger 2035 length budgets, the shipping-only configurations exhibit a similar long-term build-out.
The largest infrastructure requirements occur in configurations with access to Denmark, where captured \ce{CO2} from major source regions such as North Rhine-Westphalia must be transported over comparatively long distances before reaching a sequestration site.
In these cases, infrastructure requirements reach up to \SI{2.2}{Mt/h\cdot km} in 2035 and \SI{9.4}{Mt/h\cdot km} by 2050.

Across all topology variants, total \ce{CO2} pipeline infrastructure requirements in 2050 range from \SIrange{7.1}{9.4}{Mt/h\cdot km}, while German \ce{CO2} capture remains constant at approximately \SI{104}{Mt/a} in the climate-neutral system.
The variation in infrastructure requirements is therefore not driven by differences in captured \ce{CO2} volumes, but primarily by the spatial configuration inherited from the 2035 network.
This indicates lower average network utilization, as the additional pipeline segments primarily extend the system towards more distant sinks or connect peripheral source regions with comparatively small capture volumes.

\begin{figure}[htbp]
    \centering
    \includegraphics[width=0.5\textwidth]{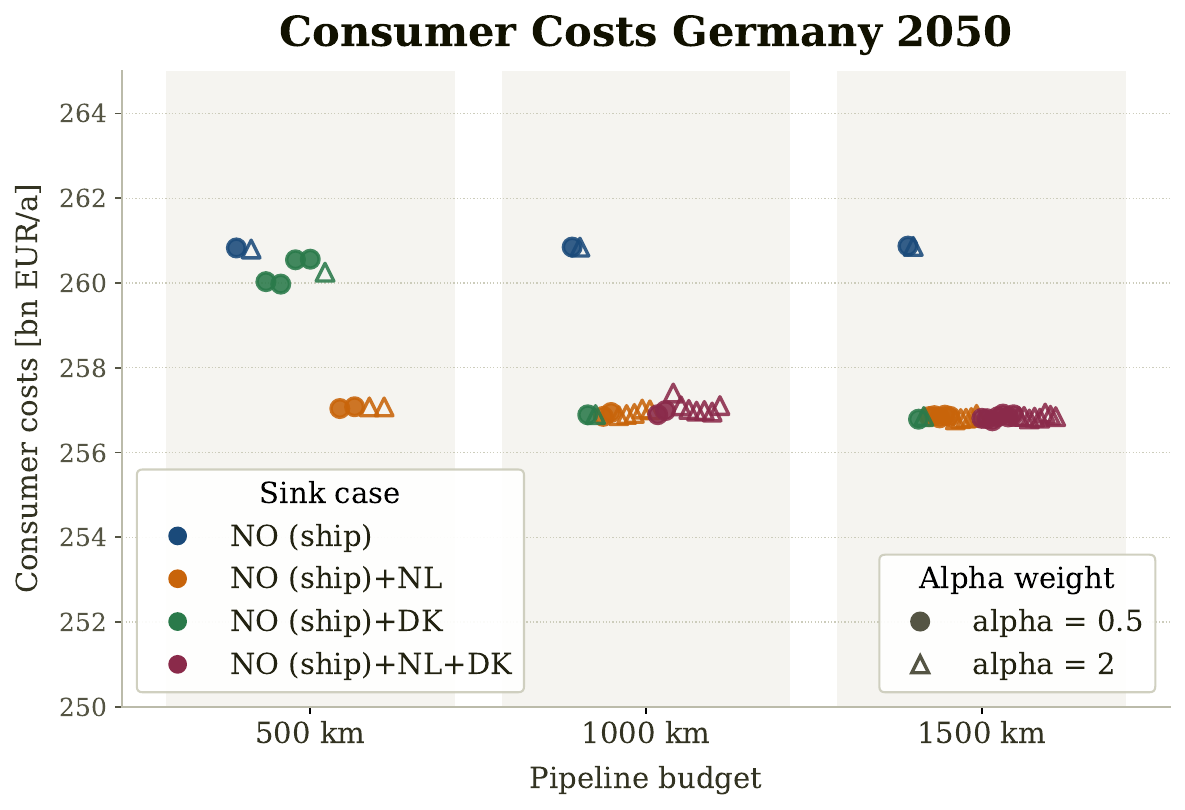}
    \caption{German consumer costs in 2050. 2035 pipeline capacities are maintained and further expansion allowed along sequestration in the German North Sea being available now.}
    \label{fig:2050_costs}
\end{figure}

German consumer costs are shown in Fig.~\ref{fig:2050_costs}. The highest consumer costs are observed for topologies with domestic sink options only or tight length limit which coincides with lower expansion and utilization of \ce{CO2} pipelines.
In particular, topologies restricted to domestic sink options exhibit consistently higher costs than scenarios with access to Dutch sequestration routes.
This finding highlights the importance of establishing interconnectors early on.

The largest long-term cost reductions are observed for topologies that connect the industrial source regions in North Rhine-Westphalia to Dutch sink infrastructure already in 2035.
For these configurations, German consumer costs are reduced by up to \SI{4.1}{bnEUR/a}.
By contrast, scenarios with only limited early pipeline deployment, such as the shipping-only case or the Danish sink configuration under tight length constraints, do not achieve comparable reductions.
This indicates that the location and connectivity of early \ce{CO2} infrastructure are more important than the total length of the initial network.
Once major industrial source regions and suitable sink corridors are connected, additional network length provides only limited long-term benefits.

In comparison, a scenario without early \ce{CO2} infrastructure leads to infrastructure requirements of \SI{2.3}{Mt/h \cdot km} for interconnectors and \SI{4.9}{Mt/h \cdot km} of domestic pipelines.
Consumer costs in Germany are below the numbers reported in Fig.\ref{fig:2050_costs} since the absence of carbon infrastructure in 2035 incentivises the earlier shift to hydrogen.
This is further discussed in the limitations and in Appendix~E.

\noindent \textbf{Additional Results}

\noindent Additional sensitivity analyses are presented in the Appendices.
Appendix~E compares the benchmark scenario without \ce{CO2} transport infrastructure to topology~002 under a \SI{500}{km} network budget.
Without a \ce{CO2} network in 2035, the energy system relies more heavily on blue hydrogen production in regions with direct access to sequestration sites, thereby utilizing a larger share of the available sequestration potential.
At the same time, the shadow price of the carbon constraint increases substantially, indicating that the remaining emissions become more costly to abate.

Appendix~F further investigates the impact of doubling the available sequestration potential in 2035 to \SI{200}{Mt/a} and allowing permanent sequestration in the German North Sea.
Increasing the available storage potential leads to higher \ce{CO2} capture and sequestration volumes while reducing overall German consumer costs.
Under these conditions, the importance of connecting high-potential industrial regions, particularly North Rhine-Westphalia via the Netherlands, becomes apparent already in 2035 for short pipeline budgets.
Allowing domestic sequestration in the German North Sea largely makes Germany self-sufficient with respect to permanent storage.
For larger network budgets, captured and sequestered volumes become relatively insensitive to the exact topology, while in several scenarios Germany becomes a net importer of \ce{CO2} from neighbouring countries for permanent offshore sequestration.

\section{Discussion and Conclusion}

Our study investigates near-term CO\textsubscript{2} transport infrastructure for Germany by generating and evaluating 60 graph-based network topologies within a large-scale integrated energy system model.
Independent of the total network length and the selected sink configuration, the introduction of a CO\textsubscript{2} transport network reduces German consumer costs in 2035 between \SIrange{22.0}{22.4}{bnEUR/a} compared to a scenario without pipeline infrastructure.

The generated topologies consistently identify transport corridors connecting North Rhine-Westphalia with the Dutch sink region.
This agrees with other climate target studies \cite{european_commission_joint_research_centre_shaping_2024, anvari_carbon_2026} and the list of projects of common and mutual interest of the European Union \cite{european_commission_commission_2025}.

In 2035, the availability of a \ce{CO2} transport infrastructure is more important than its excact topology.
German consumer costs vary only slightly across the candidate networks, irrespective of the maximum pipeline length or the available interconnector.

The differences between topology variants become more pronounced in 2050.
Early access to neighbouring sink regions, particularly through the Netherlands, reduces long-term infrastructure requirements and provides substantial system value under climate neutrality.
If sequestration is only available via shipping to the \textit{Northern Lights} project in Norway, German consumer costs in 2050 increase by up to \SI{4.1}{bnEUR/a}
Notably, a \SI{500}{km} topology connecting North Rhine-Westphalia to the Dutch sink region is sufficient to achieve favourable long-term outcomes without requiring an extensive initial network.

The resulting network focuses on highly utilized pipeline segments that connect regions with large \ce{CO2} point-source potentials and exploit the shortest available transport route to the Dutch interconnector.
Across all topologies, the \ce{CO2} transport network is primarily utilized by industrial point sources, particularly process emissions, cement production and biomass-based carbon dioxide removal.
These volumes are complemented by comparatively small contributions from waste incineration capacities.

Consequently, the analysis suggests that CCS should primarily be understood as an option for residual and hard-to-abate emissions rather than as a means to prolong fossil fuel use.
At the same time, the availability of \ce{CO2} transport and storage infrastructure may create incentives to retain fossil-based production pathways if its role is not clearly limited by policy design.
This is particularly relevant because geological sequestration potential is ultimately finite and should therefore be prioritized for applications with limited mitigation alternatives.

\noindent\textbf{Limitations}

\noindent Several limitations should be considered when interpreting the results.
The graph-theoretic topology generation strongly favours corridors connecting North Rhine-Westphalia to the Netherlands.
This result is partly driven by the predefined set of available sinks, which includes the Netherlands, Denmark and shipping-based access to the \textit{Northern Lights} project in Norway.
Alternative sink options such as Belgium or future onshore sequestration sites are not considered.

The proposed methodology remains a simplification of the underlying infrastructure planning problem.
While the graph-based approach introduces discrete topology choices that are absent in standard linear energy system models, pipeline expansion within the optimization remains continuous.
More detailed infrastructure planning approaches could represent discrete pipeline diameters, diameter-dependent specific investment costs and start-up decisions through mixed-integer formulations.

The analysis relies on a myopic optimization framework.
Infrastructure decisions made in 2035 are optimized without anticipating the goal of climate-neutrality in 2050.
This limitation becomes visible when comparing the topology scenarios to a scenario without \ce{CO2} pipelines.
The latter performs better under 2050 climate-neutral conditions because the absence of \ce{CO2} transport infrastructure in 2035 forces earlier electrification and fuel switching.
Consequently, the long-term value of early \ce{CO2} infrastructure investments cannot be assessed conclusively without a perfect-foresight framework.
A more detailed discussion of these findings is given in Appendix~E.

The representation of carbon capture is also simplified.
While the model distinguishes between major industrial sectors, it does not fully capture the heterogeneity of individual capture sources and site-specific capture costs.
By aggregating potential carbon capture sources within comparatively large spatial regions, the model neglects local collection infrastructure and assumes that all point sources within a region can be connected equally.
It therefore does not capture economies of scale associated with capture deployment, for example the higher specific costs of equipping small industrial units or combined heat and power plants with carbon capture \cite{kumar_plant_2023}.
This contributes to the comparatively flat optimum observed across different network topologies and may underestimate real-world differences between infrastructure configurations.

The analysis is further limited by the exclusive representation of pipeline transport.
In practice, early \ce{CO2} transport systems are likely to rely on combinations of trucks, rail, shipping and pipelines.
Such multimodal transport options may provide transitional solutions before dedicated pipeline infrastructure becomes available.

More generally, the suitability of energy system models for detailed infrastructure planning remains limited.
Infrastructure development is influenced by factors that are typically not represented explicitly, including public acceptance of \ce{CO2} transport infrastructure, regulatory processes, strategic industrial policy objectives and the availability of existing infrastructure corridors that could reduce permitting effort and construction costs.

Finally, the relatively coarse spatial resolution of the energy system model requires substantial aggregation of both industrial emitters and infrastructure corridors.
As a result, the identified transport corridors should be interpreted as indicative infrastructure backbones rather than detailed routing recommendations.
Similar to conventional energy system modelling approaches, the method is therefore better suited to identifying robust transport regions and recurring corridor structures than to determining exact pipeline routes.

In addition to the methodological limitations discussed above, substantial uncertainty remains regarding the future development of carbon management systems in Europe.
The availability, timing and scale of geological \ce{CO2} storage sites remain uncertain, particularly with respect to permitting processes, public acceptance and the successful realization of currently planned storage projects.
We included two sensitivities regarding the overall volume of \ce{CO2} sequestration and the availability of the German North Sea sites in Appendix~F.
The interconnectors to the Netherlands and Denmark along the shipping to Norway were chosen due to their status as projects of common and mutual interest \cite{european_commission_commission_2025}.
However, future sink availability may differ substantially from the assumptions adopted in this study.

The analysis further assumes that Germany largely retains its current energy-intensive industry and successfully transitions existing production facilities towards low-carbon production routes.
This assumption is consistent with current industrial policy objectives but remains highly uncertain under the pressure of the renewables pull \cite{seibold_balancing_2025}.
The relocation of subsectors to regions with more favorable renewable energy resources could substantially alter future \ce{CO2} transport requirements.
Such developments would affect both the spatial distribution of emissions and the relative attractiveness of different transport corridors since industrial demand is one of the primary drivers of both \ce{CO2} capture volumes and transport infrastructure requirements.
The robust corridors identified in this study should therefore be interpreted as conditional on the assumed industrial development pathways rather than as deterministic infrastructure recommendations.

\noindent\textbf{Supplementary information}

A dataset of the model results will be made available on zenodo after peer-review. The code to reproduce the experiments is available at \url{https://github.com/toniseibold/pypsa-de/tree/co2_network_v2}.

\noindent\textbf{Acknowledgements}

T.S. gratefully acknowledges funding from the Kopernikus-Ariadne project by the Federal Ministry of Research, Technology and Space (Bundesministerium für Forschung, Technik und Raumfahrt, BMFTR), grant number 03SFK5R0-2. We thank Fabian Neumann, Bobby Xiong, Eva Herrmann and many
others for useful discussions. The responsibility for the contents lies with the authors.

\noindent\textbf{Author contributions: CRediT}

T.S.: Conceptualization - Data curation - Formal Analysis - Investigation - Methodology - Software - Validation - Visualization - Writing - original draft
L.L.: Investigation - Methodology - Validation - Writing - review \& editing
T.B.: Conceptualization - Formal Analysis - Funding acquisition - Investigation - Methodology - Project administration - Supervision - Validation - Writing - review \& editing

\section*{Declarations}

The authors declare no competing interests.

\noindent\textbf{Declaration of generative AI and AI-assisted technologies in the writing process}

During the preparation of this work the author(s) used ChatGPT in order to improve wording.
After using this tool/service, the author(s) reviewed and edited the content as needed and take(s) full
responsibility for the content of the published article.
During the implementation of the graph-theoretic topology-generation workflow, GitHub Copilot was used as an AI-assisted programming tool.
All generated code was reviewed, tested and modified by the authors, who take full responsibility for the implementation and the reported results.

\newpage

\bibliographystyle{IEEEtran}
\bibliography{references}

\clearpage

\appendix

\section{PyPSA-DE} \label{app:pypsa-de}
The model follows a fully sector-coupled approach and represents energy and feedstock demands across five sectors: electricity, heating, transport, industry and agriculture.
Electricity demand can be met by both renewable and dispatchable generation technologies.
Renewable options comprise solar photovoltaics, onshore and offshore wind, while dispatchable generation includes coal, lignite, nuclear, open-cycle gas turbines, combined-cycle gas turbines, combined heat and power (CHP) plants and waste incineration facilities.

Heat demand in the building sector is supplied by a portfolio of technologies, including biomass, gas and oil boilers as well as electric heat pumps.
Buildings located in densely populated areas are assumed to be partially connected to district heating networks, allowing for centralised heat supply.
CHP plants can feed into these networks coupling the electricity and heating sector.

The transport sector represents passenger and freight transport via road and rail and additionally accounts for energy demand in shipping and aviation.
The transition from internal combustion engine vehicles to battery electric vehicles is exogenously set via a trajectory.
For waterborne transport, a shift from diesel to methanol is assumed exogenously.
Jet fuel demand can be met by both fossil and renewable pathways, with renewable jet fuel produced via biomass- and hydrogen-based processes.
The allocation between fossil and renewable jet fuel is determined endogenously within the optimisation.

The agriculture sector is represented in an aggregated manner.
An exogenously defined load captures electricity and heat demand, while oil demand for agricultural machinery is included separately to reflect limited electrification options in this sector.

The documentation of the model provides further details on modeling choices and implementation \cite{brown_pypsa-eur_2024}.

\begin{figure*}[hbtp]
    \centering
    \includegraphics[width=0.7\textwidth]{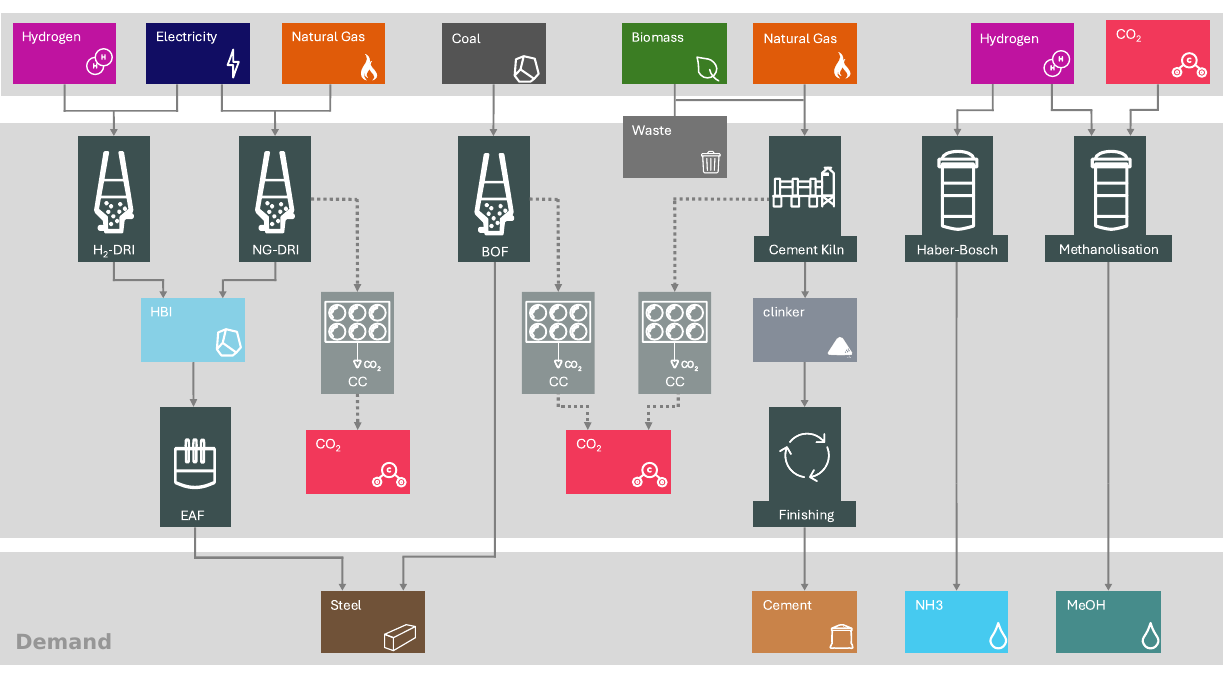}
    \caption{Overview of industrial processes and associated \ce{CO2} flows represented in the model.}
    \label{fig:industry_processes}
\end{figure*}

To assess the consistency of the industrial plant database, aggregated installed capacities are compared to historical production volumes from the JRC IDEES database \cite{jrc_database}.
In general, the reported capacities are consistent with historical production levels, although cement production exhibits noticeable overcapacity.
In cases where installed capacities are insufficient to satisfy demand, the model can endogenously expand industrial production capacity.

\begin{figure}[hbt]
    \centering
    \includegraphics[width=1.0\linewidth]{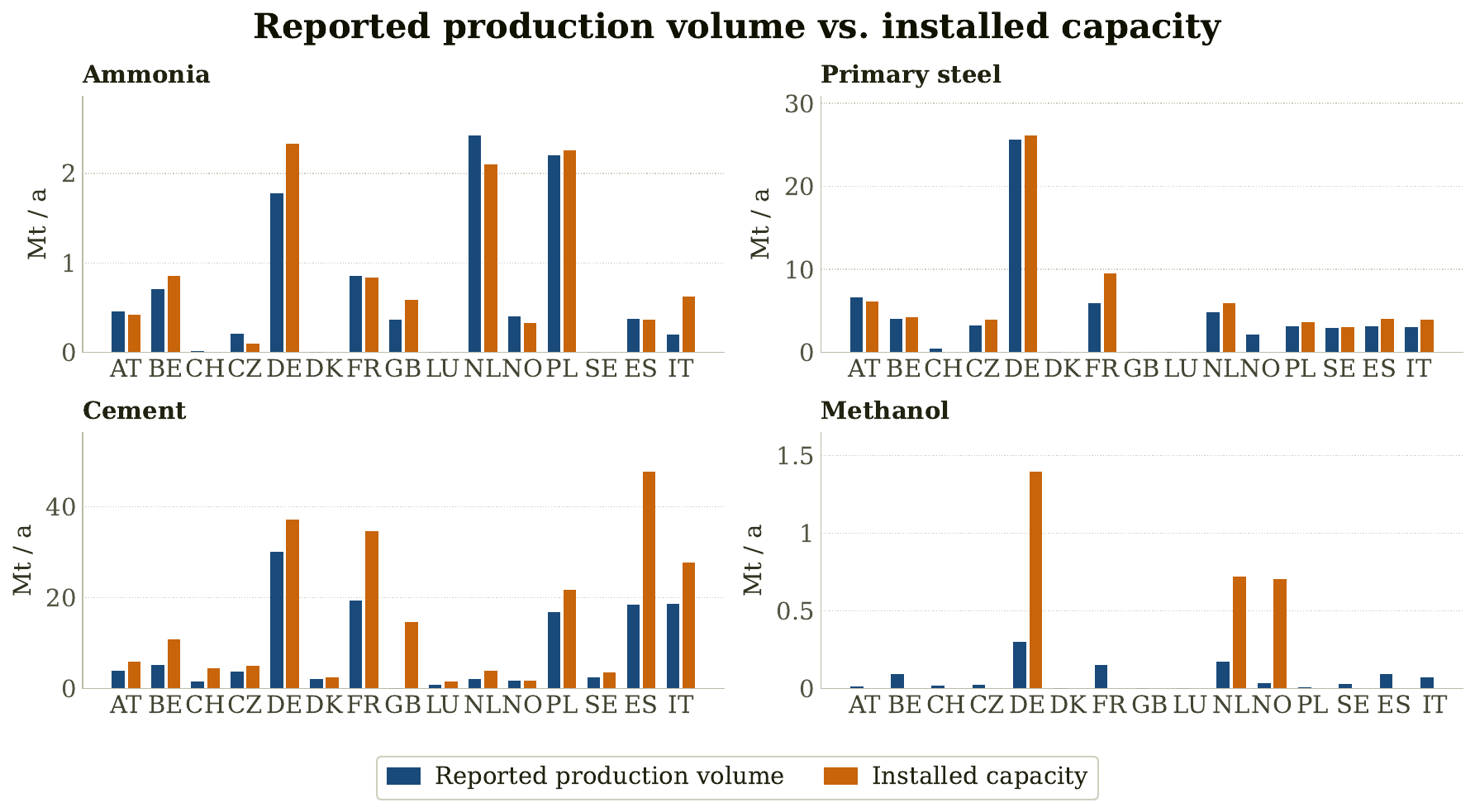}
    \caption{Aggregated installed production capacities for primary steel,
    cement, ammonia and methanol production (orange)
    \cite{global_energy_monitor_global_2025,
    global_energy_monitor_global_2025-1,
    neuwirth_modelling_2024}
    compared to reported production volumes in 2020 (blue)
    \cite{jrc_database}.}
    \label{fig:plants_demand}
\end{figure}

Our model focuses on Germany and represents the energy system using 89 spatial nodes, of which 23 are located in Germany.
The remaining nodes are distributed across Europe with increased spatial resolution in countries that have access to offshore \ce{CO2} storage sites, particularly in coastal regions.
The landlocked countries Austria, Switzerland and the Czech Republic are represented by a single node each.
This spatial resolution ensures that inland \ce{CO2} point sources remain geographically separated from coastal sequestration sites.
As a result, captured \ce{CO2} cannot be implicitly transported to storage locations through spatial aggregation and must instead utilize explicit transport infrastructure.

\begin{figure}
    \centering
    \includegraphics[width=0.5\linewidth]{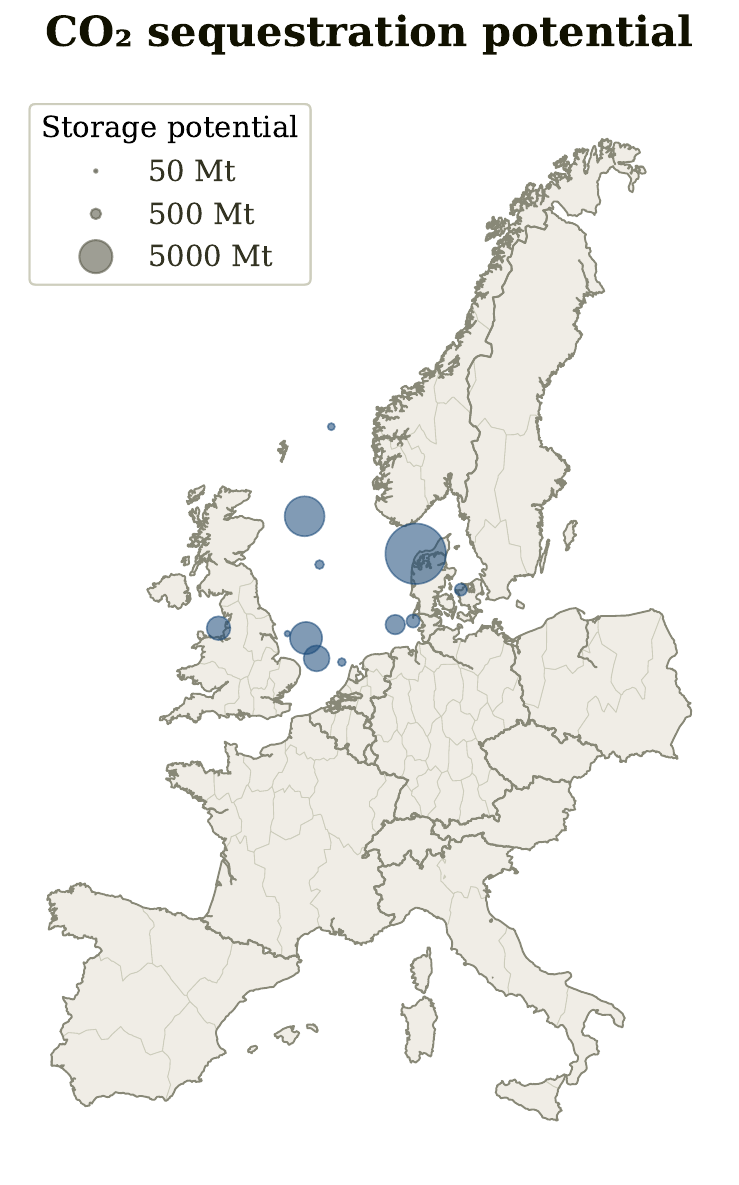}
    \caption{Model scope and sequestration potential.}
    \label{fig:scope_potential}
\end{figure}

\clearpage

\section{Techno-economic data} \label{app:techno-economic}

\begin{longtable}{@{}lllll@{}}
    \caption{Techno-economic data for \ce{CO2} transport technology.} \label{tab:co2_tech} \\
    \toprule
    technology & parameter & value & unit & source \\
    \midrule
    \ce{CO2} pipeline (\diameter \SI{70}{cm})
        & FOM & 0.9 & \%/year & \cite{dea_carbon} \\
        & overnight investment & 2858.6 & EUR/(t\textsubscript{\ce{CO2}}/h)km & \cite{schutzenhofer_machbarkeitsstudie_2024} \\
        & lifetime & 50 & years & \cite{dea_carbon} \\
    \ce{CO2} pipeline (\diameter \SI{70}{cm}, underwater) 
        & FOM & 0.9 & \%/year & \cite{dea_carbon} \\
        & overnight investment & 5717.2 & EUR/(t\textsubscript{\ce{CO2}}/h)km & \\
        & lifetime & 50 & years & \cite{dea_carbon} \\
    \ce{CO2} compression
        & FOM & 5.0 & \%/year & \cite{schutzenhofer_machbarkeitsstudie_2024}  \\
        & overnight investment & 33288 & EUR/t\textsubscript{\ce{CO2}} & \cite{schutzenhofer_machbarkeitsstudie_2024}  \\
        & lifetime & 25 & year & \\
        & electricity-input & 0.1051 & MWh\textsubscript{el}/t\textsubscript{\ce{CO2}} \\
    \ce{CO2} sequestration 
        & marginal cost & 35 & EUR/t & \cite{schutzenhofer_machbarkeitsstudie_2024}  \\
    \ce{CO2} shipping
        & marginal cost & 50 & EUR/t & \cite{malz_assessing_2025} \\
    \bottomrule
\end{longtable}

\begin{longtable}{@{}lllll@{}}
    \caption{Excerpt from \textit{technology-data} for the year 2035 of the technologies most relevant to this work \cite{technology-data} in EUR$_{2025}$. Own assumption where source is missing.} \label{tab:technology-parameters} \\
    \toprule
    technology & parameter & value & unit & source \\
    \midrule
    Electrolysis
        & FOM & 4.0 & \%/year & \cite{dea_renewable_fuels} \\
        & efficiency & 0.6374 & per unit & \cite{dea_renewable_fuels} \\
        & efficiency-heat & 0.2039 & per unit & \cite{dea_renewable_fuels} \\
        & investment & 1697.4017 & EUR/kW\textsubscript{el} & \\
        & lifetime & 25.0 & years & \cite{dea_renewable_fuels} \\
    Fischer-Tropsch
        & FOM & 6.2815 & \%/year & \cite{dea_renewable_fuels}\\
        & VOM & 4.9471 & EUR/MWh\textsubscript{FT} & \cite{dea_renewable_fuels} \\
        & capture rate & 0.9 & per unit & \cite{hannula_co-production_2015} \\
        & \ce{CO2}-input & 0.3135 & t\_CO2/MWh\textsubscript{FT} & \cite{dea_renewable_fuels} \\
        & efficiency & 0.719 & per unit &  \\
        & electricity-input & 7 & MWh\textsubscript{el}/MWh\textsubscript{FT} & \cite{dea_renewable_fuels} \\
        & hydrogen-input & 1392 & MWh\textsubscript{H\textsubscript{2}}/MWh\textsubscript{FT} & \cite{dea_renewable_fuels} \\
        & investment & 1805017 & EUR/kw\textsubscript{FT} & \cite{dea_renewable_fuels} \\
        & lifetime & 25.0 & years & \cite{dea_renewable_fuels} \\
    Methanolisation
        & FOM & 2.9787 & \%/year & \cite{dea_renewable_fuels} \\
        & capture rate & 0.9 & per unit & \cite{hannula_co-production_2015} \\
        & \ce{CO2}-input & 248 & t\textsubscript{\ce{CO2}}/MWh\textsubscript{MeOH} & \cite{bazzanella_technology_2017} \\
        & electricity-input & 271 & MWh\textsubscript{el}/MWh\textsubscript{MeOH} & \cite{bazzanella_technology_2017} \\
        & heat-output & 0.1 & MWh\textsubscript{th}/MWh\textsubscript{MeOH} & \cite{bazzanella_technology_2017} \\
        & hydrogen-input & 1138 & MWh\textsubscript{H\textsubscript{2}}/MWh\textsubscript{MeOH} & \cite{bazzanella_technology_2017} \\
        & investment & 1282.8604 & EUR/kW\textsubscript{MeOH} & \cite{dea_renewable_fuels} \\
    \bottomrule
    \newpage
    \toprule
    technology & parameter & value & unit & source \\
    \midrule
    Methanolisation
        & lifetime & 30.0 & years & \cite{dea_renewable_fuels} \\
    Biomass-to-MeOH
        & C in fuel & 0.4197 & per unit & \cite{millinger_are_2022} \\
        & C stored & 0.5803 & per unit & \cite{millinger_are_2022} \\
        & CO2 stored & 0.2128 & t\textsubscript{\ce{CO2}}/MWh\textsubscript{th} & \cite{millinger_are_2022} \\
        & FOM & 1.5331 & \%/year & \cite{dea_renewable_fuels} \\
        & VOM & 18.1877 & EUR/MWh\textsubscript{MeOH} & \cite{dea_renewable_fuels} \\
        & capture rate & 0.9 & per unit & \cite{hannula_co-production_2015} \\
        & efficiency & 0.62 & per unit & \cite{dea_renewable_fuels} \\
        & efficiency-electricity & 0.02 & MWh\textsubscript{el}/MWh\textsubscript{th} & \cite{dea_renewable_fuels} \\
        & efficiency-heat & 0.22 & per unit & \cite{dea_renewable_fuels} \\
        & investment & 3370.9305 & EUR/kW\textsubscript{MeOH} & \cite{dea_renewable_fuels} \\
        & lifetime & 20.0 & years & \cite{dea_renewable_fuels} \\
    MeOHtO/A
        & FOM & 3.0 & \%/year &  \\
        & VOM & 40.0741 & EUR/t\textsubscript{HVC} & \cite{bazzanella_technology_2017} \\
        & \ce{CO2}-output & 0.61 & t\textsubscript{\ce{CO2}}/t\textsubscript{HVC} & \cite{bazzanella_technology_2017} \\
        & electricity-input & 1.39 & MWh\textsubscript{el}/t\textsubscript{HVC} & \cite{bazzanella_technology_2017} \\
        & investment & 3510490.6 & EUR/(t\textsubscript{HVC}/h) & \cite{bazzanella_technology_2017} \\
        & lifetime & 30.0 & years & \cite{dea_renewable_fuels} \\
        & methanol-input & 18.03 & MWh\textsubscript{MeOH}/t\textsubscript{HVC} & \cite{bazzanella_technology_2017} \\
    H\textsubscript{2} DRI 
        & FOM               & 11.3      & \%/year           & \cite{mpp} \\
        & electricity-input & 1.03      & MWh\textsubscript{el}/t\textsubscript{HBI}    & \cite{mpp} \\
        & hydrogen-input    & 2.1       & MWh\textsubscript{H\textsubscript{2}}/t\textsubscript{HBI}    & \cite{mpp} \\
        & investment        & 5378698.9 & EUR/t\textsubscript{HBI}/h      & \cite{mpp} \\
        & lifetime          & 40.0      & years             & \cite{mpp} \\
        & ore-input         & 1.59      & t\textsubscript{ore}/t\textsubscript{HBI}     & \cite{mpp} \\
    Gas DRI
        & FOM               & 11.3      & \%/year           & \cite{mpp} \\
        & electricity-input & 1.03      & MWh\textsubscript{el}/t\textsubscript{HBI}    & \cite{mpp} \\
        & gas-input    & 2.78       & MWh\textsubscript{H\textsubscript{2}}/t\textsubscript{HBI}    & \cite{mpp} \\
        & investment        & 5378698.9 & EUR/t\textsubscript{HBI}/h      & \cite{mpp} \\
        & lifetime          & 40.0      & years             & \cite{mpp} \\
        & ore-input         & 1.59      & t\textsubscript{ore}/t\textsubscript{HBI}     & \cite{mpp} \\
    EAF
        & FOM               & 30.0      & \%/year          & \cite{mpp} \\
        & electricity-input & 0.64    & MWh\textsubscript{el}/t\textsubscript{steel}  & \cite{mpp} \\
        & hbi-input         & 1.0       & t\textsubscript{HBI}/t\textsubscript{steel}   & \cite{mpp} \\
        & investment        & 2312992.7 & EUR/t\textsubscript{steel}/h    & \cite{mpp} \\
        & lifetime          & 40.0      & years             & \cite{mpp} \\
    BF-BOF
        & FOM               & 14.18      & \%/year          & \cite{mpp} \\
        & coal-input        & 5.34       & t\textsubscript{HBI}/t\textsubscript{steel}   & \cite{mpp} \\
        & ore-input         & 1.539     & t\textsubscript{ore}/t\textsubscript{steel} & \cite{mpp} \\
        & investment   & 9602774.8 & EUR/t\textsubscript{steel}/h    & \cite{mpp} \\
        & lifetime          & 40.0      & years             & \cite{mpp} \\
    Haber-Bosch
        & FOM              & 3.0      & \%/year         & \cite{dea_renewable_fuels} \\
        & VOM              & 0.03   & EUR/MWh\textsubscript{NH\textsubscript{3}}    & \cite{dea_renewable_fuels} \\
        & electricity-input& 0.25   & MWh\textsubscript{el}/MWh\textsubscript{NH\textsubscript{3}}& \cite{bazzanella_technology_2017} \\
    \bottomrule
    \newpage
    \toprule
    technology & parameter & value & unit & source \\
    \midrule
    Haber-Bosch
        & hydrogen-input   & 1.15   & MWh\textsubscript{H\textsubscript{2}}/MWh\textsubscript{NH\textsubscript{3}}& \cite{bazzanella_technology_2017} \\
        & investment       & 1675.2 & EUR/kW\textsubscript{NH\textsubscript{3}}     & \cite{bazzanella_technology_2017} \\
        & lifetime         & 30.0     & years           & \cite{dea_renewable_fuels} \\
        & nitrogen-input   & 0.16   & t\textsubscript{N\textsubscript{2}}/MWh\textsubscript{NH\textsubscript{3}}  & \cite{bazzanella_technology_2017} \\
    Air-separation
        & FOM              & 3.0      & \%/year         & \cite{dea_renewable_fuels} \\
        & electricity-input& 0.25     & MWh\textsubscript{el}/t\textsubscript{N\textsubscript{2}}& \cite{dea_renewable_fuels} \\
        & investment& 941625.8 & EUR/t\textsubscript{N\textsubscript{2}/h}     & \cite{dea_renewable_fuels} \\
        & lifetime         & 30.0     & years           & \cite{dea_renewable_fuels} \\
    Cement-kiln
        & FOM              & 4.0    & \%/year         & \cite{nijs_jrc-eu-times_2019} \\
        & VOM              & 6679   & EUR/t\textsubscript{clinker} & \cite{nijs_jrc-eu-times_2019} \\
        & electricity-input& 0.07   & MWh\textsubscript{el}/t\textsubscript{clinker}& \cite{nijs_jrc-eu-times_2019} \\
        & heat-input       & 0.94   & MWh\textsubscript{heat}/t\textsubscript{clinker}& \cite{nijs_jrc-eu-times_2019} \\
        & gas-input        & 0.0002 & MWh\textsubscript{gas}/t\textsubscript{clinker}& \cite{nijs_jrc-eu-times_2019} \\
        & investment      & 1462704.4 & EUR/t\textsubscript{clinker}/h     & \cite{nijs_jrc-eu-times_2019} \\
        & lifetime         & 30.0   & years           & \cite{nijs_jrc-eu-times_2019} \\
    Cement Heat
        & biomass-efficiency& 0.9   & MWh\textsubscript{heat}/MWh\textsubscript{biomass}& \\
        & waste-efficiency  & 0.9   & MWh\textsubscript{heat}/MWh\textsubscript{waste}  & \\
        & gas-efficiency    & 1.0   & MWh\textsubscript{heat}/MWh\textsubscript{gas}    & \\
    Cement Finishing
        & FOM              & 30.0   & \%/year         & \cite{nijs_jrc-eu-times_2019} \\
        & VOM              & 4.0    & EUR/t\textsubscript{cement}         & \cite{nijs_jrc-eu-times_2019} \\
        & clinker-input    & 0.656  & t\textsubscript{clinker}/t\textsubscript{cement} & \cite{nijs_jrc-eu-times_2019} \\
        & electricity-input& 0.17   & MWh\textsubscript{el}/t\textsubscript{cement}& \cite{nijs_jrc-eu-times_2019} \\
        & investment      & 117016.4 & EUR/t\textsubscript{cement}/h     & \cite{nijs_jrc-eu-times_2019} \\
        & lifetime         & 30.0   & years           & \cite{nijs_jrc-eu-times_2019} \\
    Industry Post-CC
        & FOM              & 3.0    & \%/year         & \cite{dea_carbon} \\
        & capture rate     & 0.925  & per unit        & \cite{dea_carbon} \\
        & electricity-input& 0.029  & MWh\textsubscript{el}/t\textsubscript{\ce{CO2}}& \cite{dea_carbon} \\
        & investment       & 3017603.0 & EUR/t\textsubscript{\ce{CO2}}/h     & \cite{dea_carbon} \\
        & lifetime         & 25.0   & years           & \cite{dea_carbon} \\
    Gas CHP CC
        & investment       & 5829460.4 & EUR/t\textsubscript{\ce{CO2}}/h    & \cite{dea_carbon} \\
    Biomass CHP CC
        & investment       & 3206203.2 & EUR/t\textsubscript{\ce{CO2}}/h    & \cite{dea_carbon} \\
    Waste CHP CC
        & investment       & 3497676.2 & EUR/t\textsubscript{\ce{CO2}}/h    & \cite{dea_carbon} \\
    Gas PP/CHP CC retrofit
        & investment       & 6035206.0 & EUR/t\textsubscript{\ce{CO2}}/h    &  \\
    Biomass PP/CHP CC retrofit
        & investment       & 3319363.3 & EUR/t\textsubscript{\ce{CO2}}/h    & \cite{dea_carbon} \\
    Waste PP/CHP CC retrofit
        & investment       & 3621123.6 & EUR/t\textsubscript{\ce{CO2}}/h    &  \\
    \bottomrule
\end{longtable}

We assume that there is enough waste heat with industry processes to supply a post combustion carbon capture process to recover the sorbent. The post-combustion investment costs depend on the concentration of \ce{CO2} in the flue gas.
The higher the concentration of \ce{CO2} in the flue gas the smaller the unit necessary. With the factors presented in Table~\ref{tab:post_cc}, we derive different investment costs for industry CC and post combustion CC (retrofit) of power and CHP plants.

\begin{table}[ht]
    \begin{center}
    \caption{Post combustion flue gas concentration and CAPEX factor taken from \cite{dea_carbon}.}
    \label{tab:post_cc}
    \begin{tabular}{lcc}
        \toprule
        technology & vol\% & CAPEX factor \\
        \midrule
        gas power plant & $2-4$ & 2 \\
        waste power plant & $9-14$ & 1.2 \\
        biomass power plant & $13-17$ & 1.1 \\
        cement plant & $20-30$ & 1.0 \\
    \end{tabular}
\end{center}
\end{table}

\newpage

\section{Individal topologies} \label{app:topo}

\begin{figure}[htbp]
    \centering
    \includegraphics[width=0.75
    \textwidth]{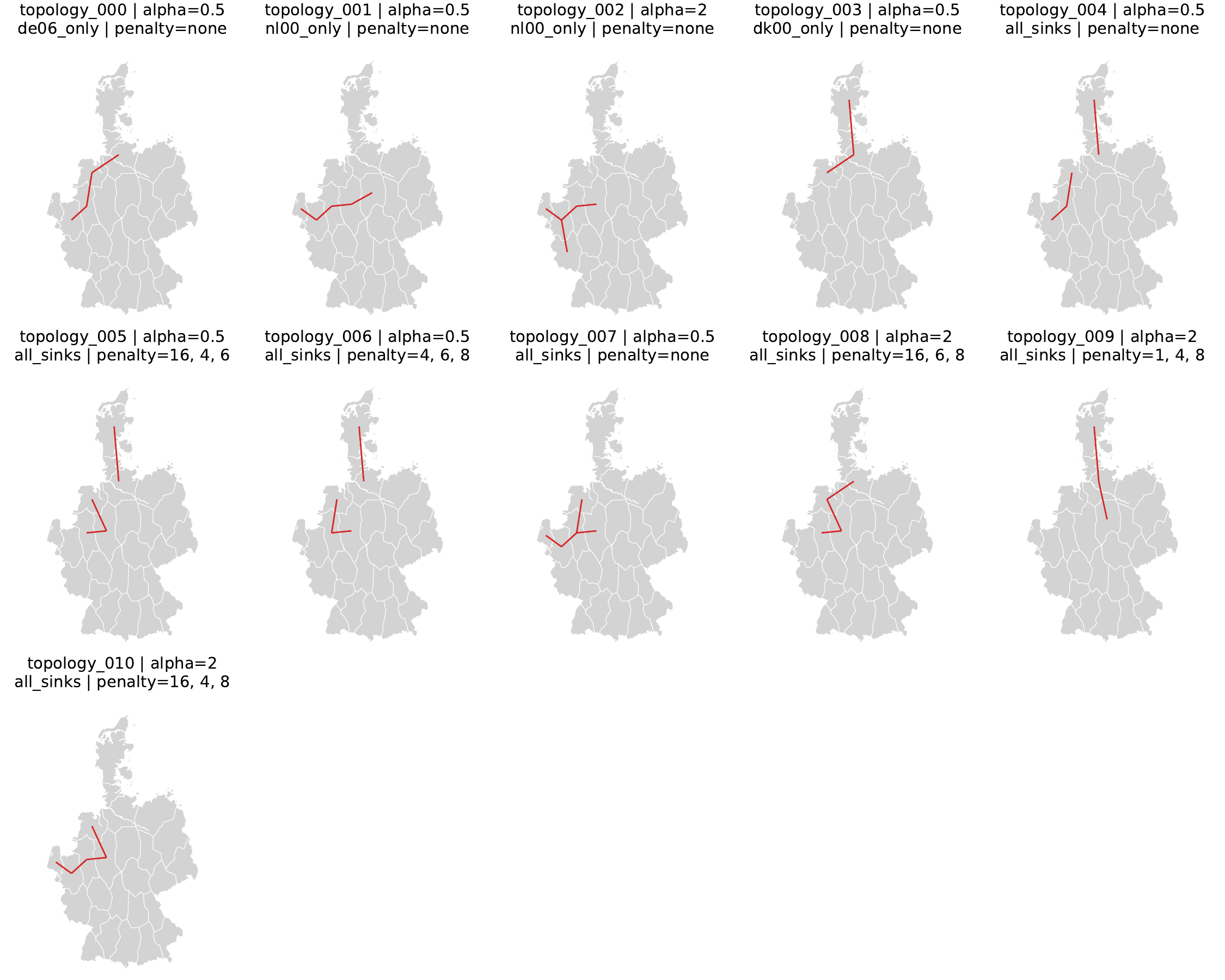}
    \caption{Topologies generated with a maximum length of 500km.}
\end{figure}

\begin{figure}[htbp]
    \centering
    \includegraphics[width=0.75\textwidth]{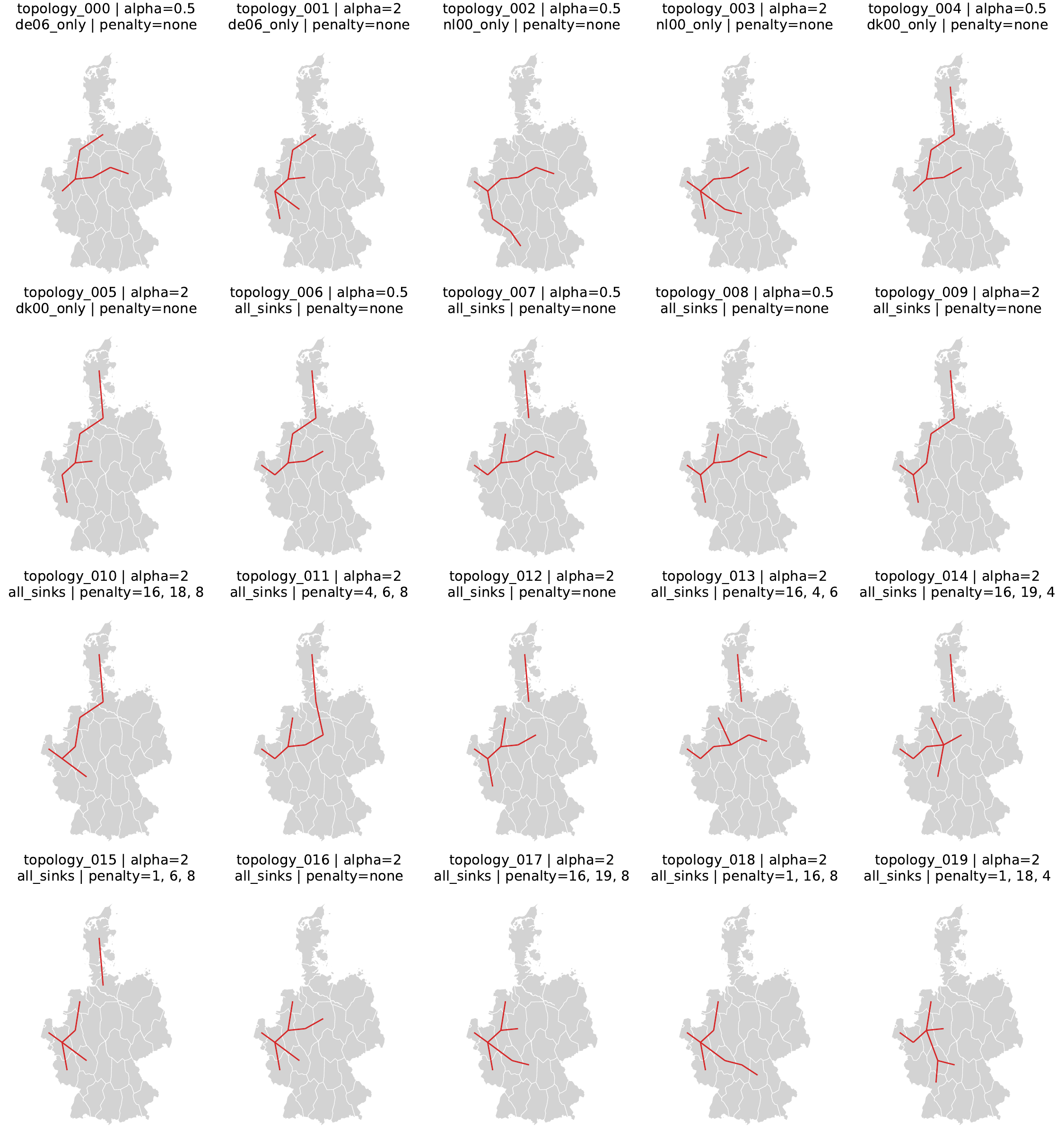}
    \caption{Topologies generated with a maximum length of 1000km.}
\end{figure}

\begin{figure}[htbp]
    \centering
    \includegraphics[width=0.75\textwidth]{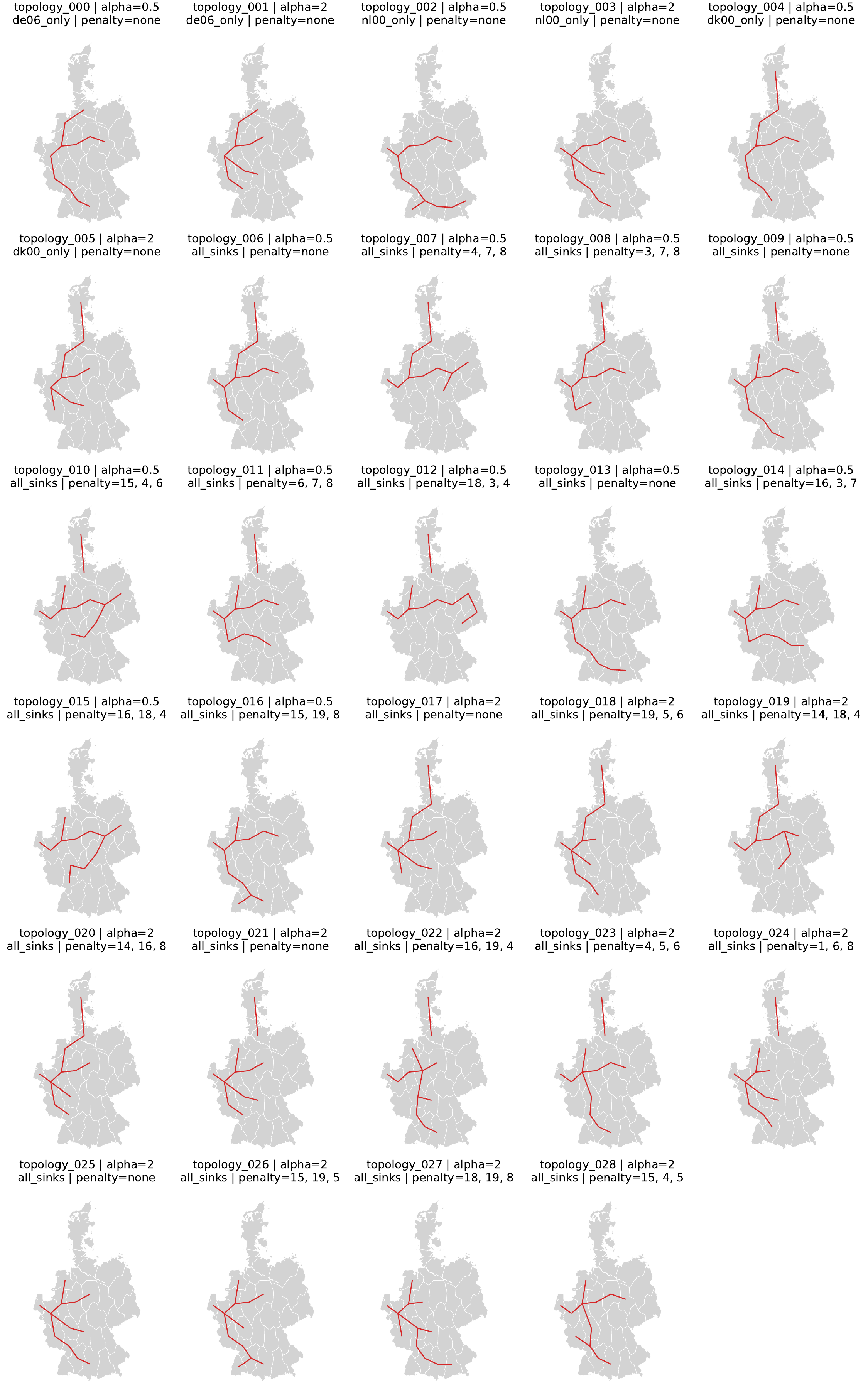}
    \caption{Topologies generated with a maximum length of 1500km.}
\end{figure}

\newpage

\section{German consumer cost savings by category}\label{app:cc_category}

We find that carbon infrastructure saves \SIrange{22.0}{22.4}{bnEUR/a} in 2035 compared to a scenario without \ce{CO2} pipelines. Fig.~\ref{app:cc_category} shows the savings compared to that scenario for every topology run.

\begin{figure}[htbp]
    \centering
    \includegraphics[width=0.8\textwidth]{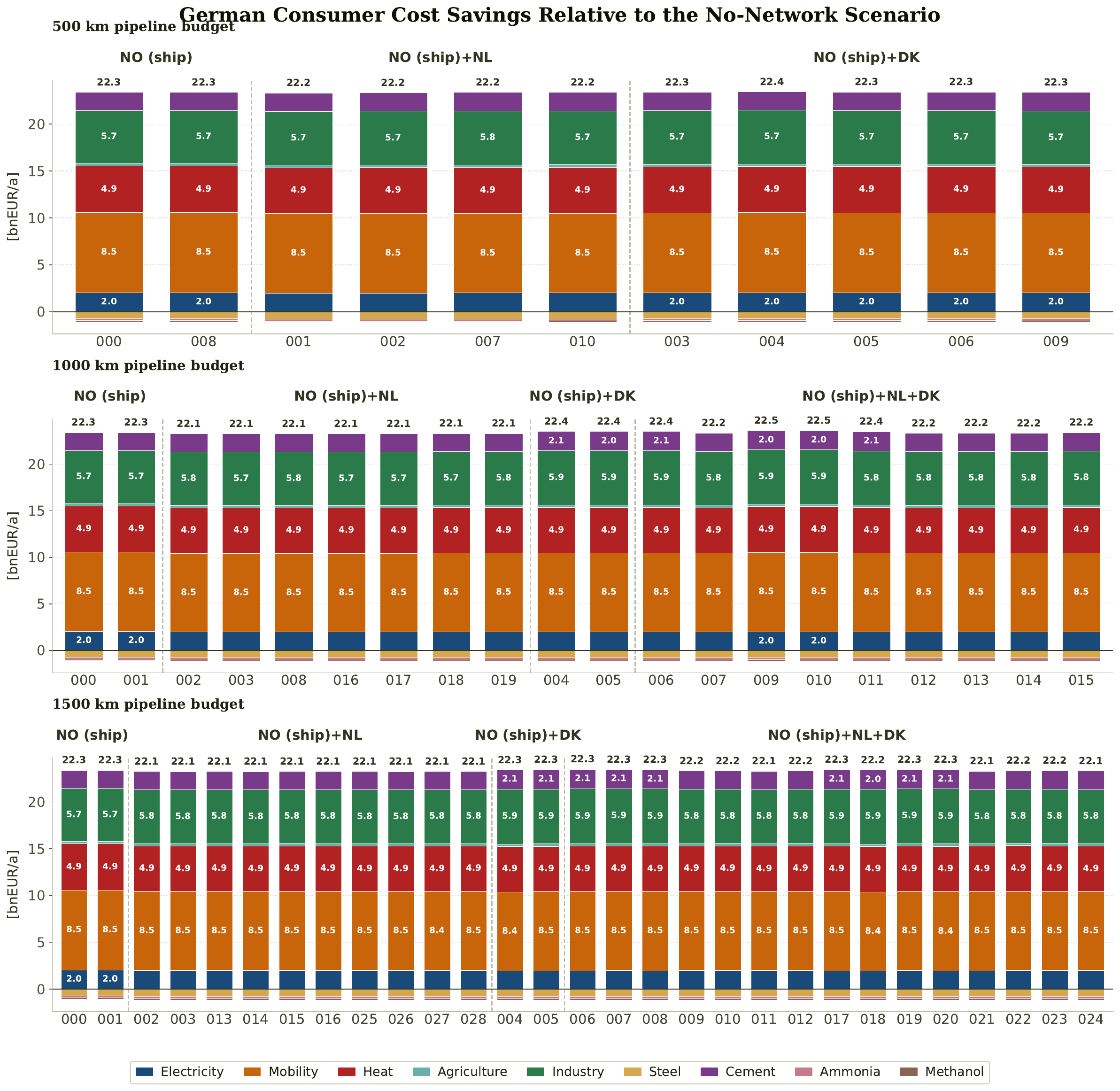}
    \caption{German consumer cost savings in 2035 compared to a scenario without \ce{CO2} pipelines by category.}
    \label{app:cc_category}
\end{figure}

\section{No \ce{CO2} pipeline scenario}\label{app:comparison}

To compare the topology scenarios against a baseline, we introduced a scenario in which no \ce{CO2} pipeline infrastructure (\textit{No \ce{CO2} network 2035}) is available in 2035 in the main part of the paper.
When extending our analysis to 2050, this scenario optimizes with the investment decisions from 2035 but is then additionally allowed to expand \ce{CO2} infrastructure unlimited.

\noindent\textbf{Carbon Balance}

\begin{figure}[htbp]
    \centering
    \includegraphics[width=0.6\linewidth]{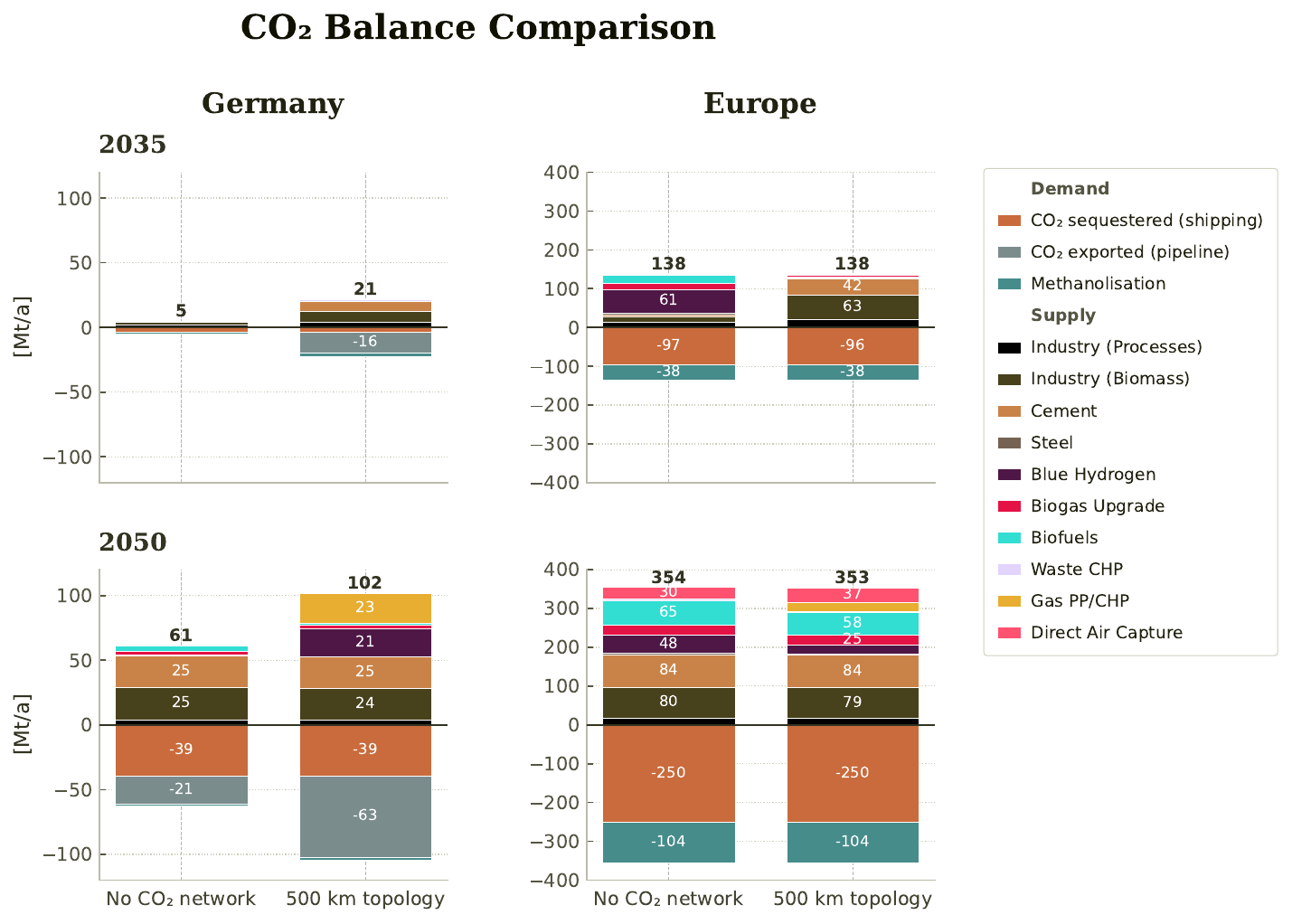}
    \caption{CO$_2$ balance for Germany (left) and Europe (right) for a scenario without carbon infrastructure and a \SI{500}{km} network (topology 002). On the top the year 2035 and on the bottom the year 2050.}
    \label{comp:co2_balance}
\end{figure}

\noindent Fig.~\ref{comp:co2_balance} shows the carbon balances for 2035 (top) and 2050 (bottom) on a German (left) and European level (right).
The absence of transport infrastructure strongly affects the spatial distribution of carbon capture.
In Germany, only around \SI{5}{Mt/a} of \ce{CO2} are captured in the \textit{No \ce{CO2} network 2035} scenario, compared to \SI{21}{Mt/a} in topology 002, which provides a \SI{500}{km} network connecting North Rhine-Westphalia to the Netherlands.

At the European level total captured \ce{CO2} volumes remain comparatively stable at around \SI{138}{Mt/a}.
Rather than reducing overall capture, the absence of infrastructure shifts capture towards regions with direct access to sequestration sites.
This effect is particularly visible for spatially flexible processes such as blue hydrogen production, biogas upgrading and biofuel production.
In contrast, industrial point sources remain fixed in their location and therefore cannot relocate to regions with easier access to storage infrastructure.

A similar pattern can be observed in 2050.
Germany captures only \SI{61}{Mt/a} of \ce{CO2} in the \textit{No \ce{CO2} network 2035} scenario compared to \SI{102}{Mt/a} when \ce{CO2} infrastructure is already available in 2035.
The reduction is primarily associated with lower capture volumes from blue hydrogen production and backup power generation.
At the European scale, blue hydrogen production remains more prominent in the scenario without early \ce{CO2} infrastructure.

The results further indicate a shift in the composition of carbon management technologies.
In the absence of early \ce{CO2} infrastructure, the sequestration capacity that would otherwise be utilized by blue hydrogen production is instead used by direct air capture and gas-fired backup capacities.
This suggests that under climate-neutrality constraints, these technologies may provide a more efficient use of limited sequestration resources than producing hydrogen from fossil fuels with carbon capture.

\noindent\textbf{Hydrogen Balance}

\begin{figure}
    \centering
    \includegraphics[width=0.6\linewidth]{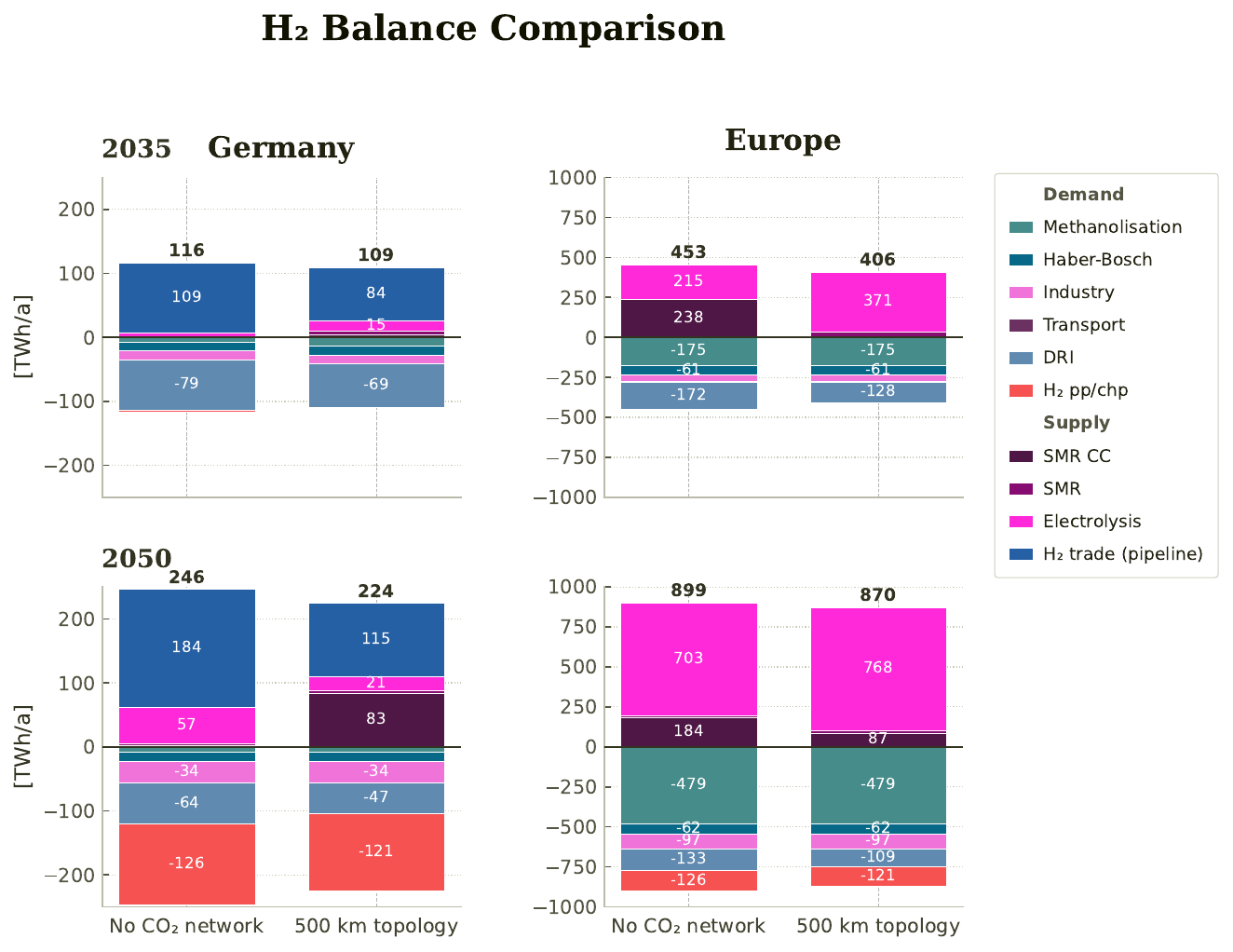}
    \caption{2035 H$_2$ balance for Germany (left) and Europe (right) between a scenario without carbon infrastructure and a \SI{500}{km} network (topology 002).}
    \label{comp:h2_bal}
\end{figure}

\noindent The \textit{No \ce{CO2} network 2035} also has influence on the hydrogen balances.
Fig.~\ref{comp:h2_bal} shows the hydrogen balance for Germany (left) and Europe (right) for the years 2035 (top) and 2050 (bottom).
In Germany, hydrogen use increases by \SI{7}{TWh/a} with the additional hydrogen mainly supplied through imports from European partners.
By contrast, the availability of \ce{CO2} infrastructure allows for higher domestic hydrogen production, including both electrolysis and SMR CC.
At the European level, the absence of \ce{CO2} infrastructure is associated with \SI{47}{TWh/a} higher hydrogen use, corresponding to about \SI{11.6}{\percent} of total hydrogen demand.

Roughly half of the hydrogen is supplied by blue hydrogen production.
This suggests that SMR CC gives the system additional flexibility compared to relying exclusively on electrolysis.
The additional hydrogen is primarily used in primary steel production, leading to a faster transformation of this industrial sector.
A similar but weaker effect is observed in 2050.
German hydrogen demand is \SI{22}{TWh/a} higher in the scenario without early \ce{CO2} infrastructure, and demand is largely covered by imports from European partners.
In contrast, scenarios with \ce{CO2} infrastructure available in 2035 allow for some domestic blue hydrogen production in Germany in 2050.
Lower hydrogen demand in these scenarios is linked to lower use of hydrogen-based direct reduction in primary steel production.
Instead, the system can rely more strongly on gas-based direct reduction with carbon capture where this is cost-effective.
This does not imply a technological lock-in at the plant level, since the model allows switching between gas and hydrogen within the same direct reduction plant and only the carbon capture capacity remains part of the brownfield assets.
At the European level, hydrogen use in 2050 is \SI{29}{TWh/a} higher in the \textit{No \ce{CO2} network 2035} scenario, corresponding to approximately \SI{3}{\percent} of total hydrogen demand.
This additional hydrogen is mainly used for hydrogen-based direct reduction in the steel sector.

Compared to 2035, the role of blue hydrogen is reversed.
Because blue hydrogen production capacities built in the transition period are inherited, they continue to produce around \SI{180}{TWh/a} in 2050 in the scenario without \ce{CO2} pipeline infrastructure in 2035.

\noindent\textbf{Carbon Map}

\begin{figure}[htbp]
    \centering
    \includegraphics[width=0.6\linewidth]{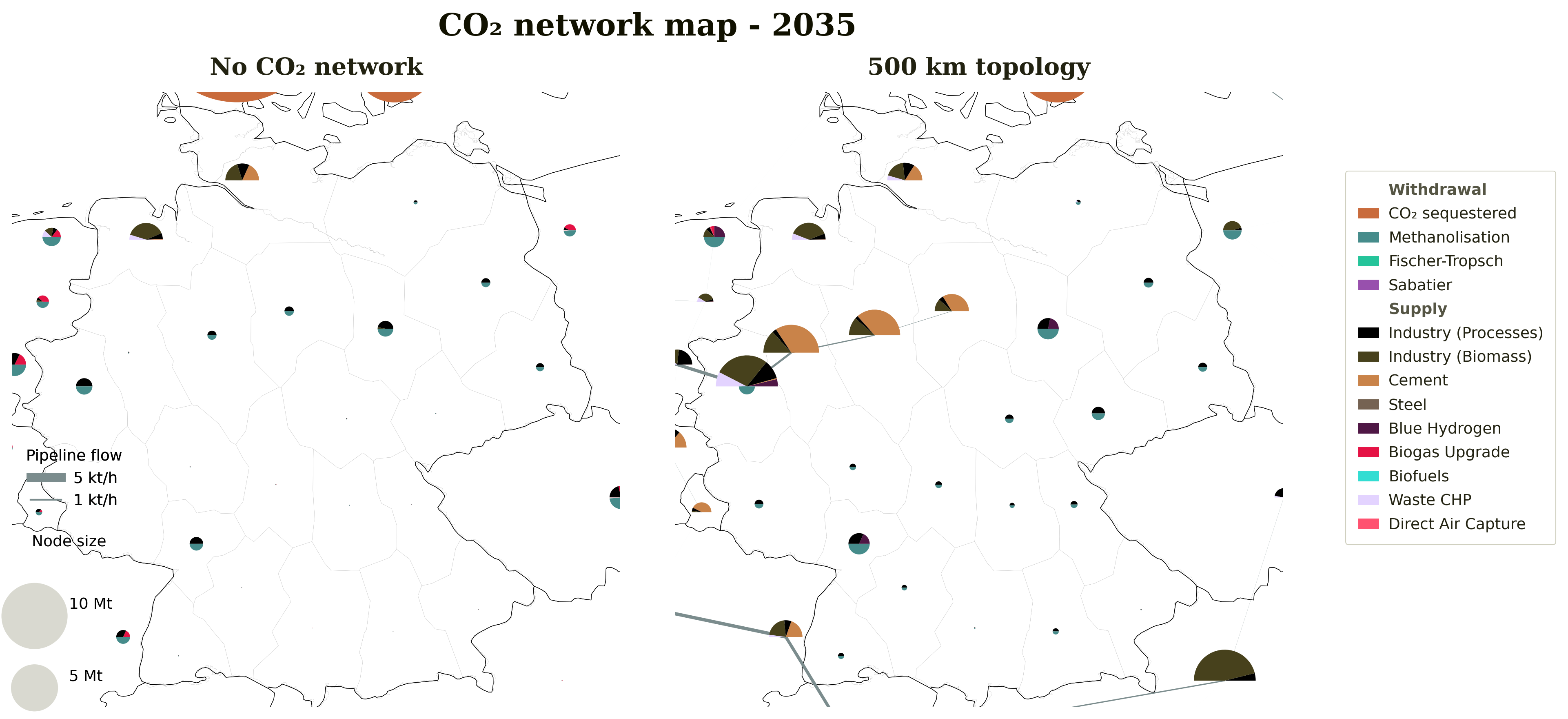}
    \caption{Carbon map for a scenario without carbon infrastructure in 2035 (left) and a \SI{500}{km} network (topology 002). Lines represent \ce{CO2} pipelines, pie charts represent \ce{CO2} captured (upper half) and utilized/sequestered (lower half).}
    \label{comp:co2_map_2035}
\end{figure}

\begin{figure}[htbp]
    \centering
    \includegraphics[width=0.6\linewidth]{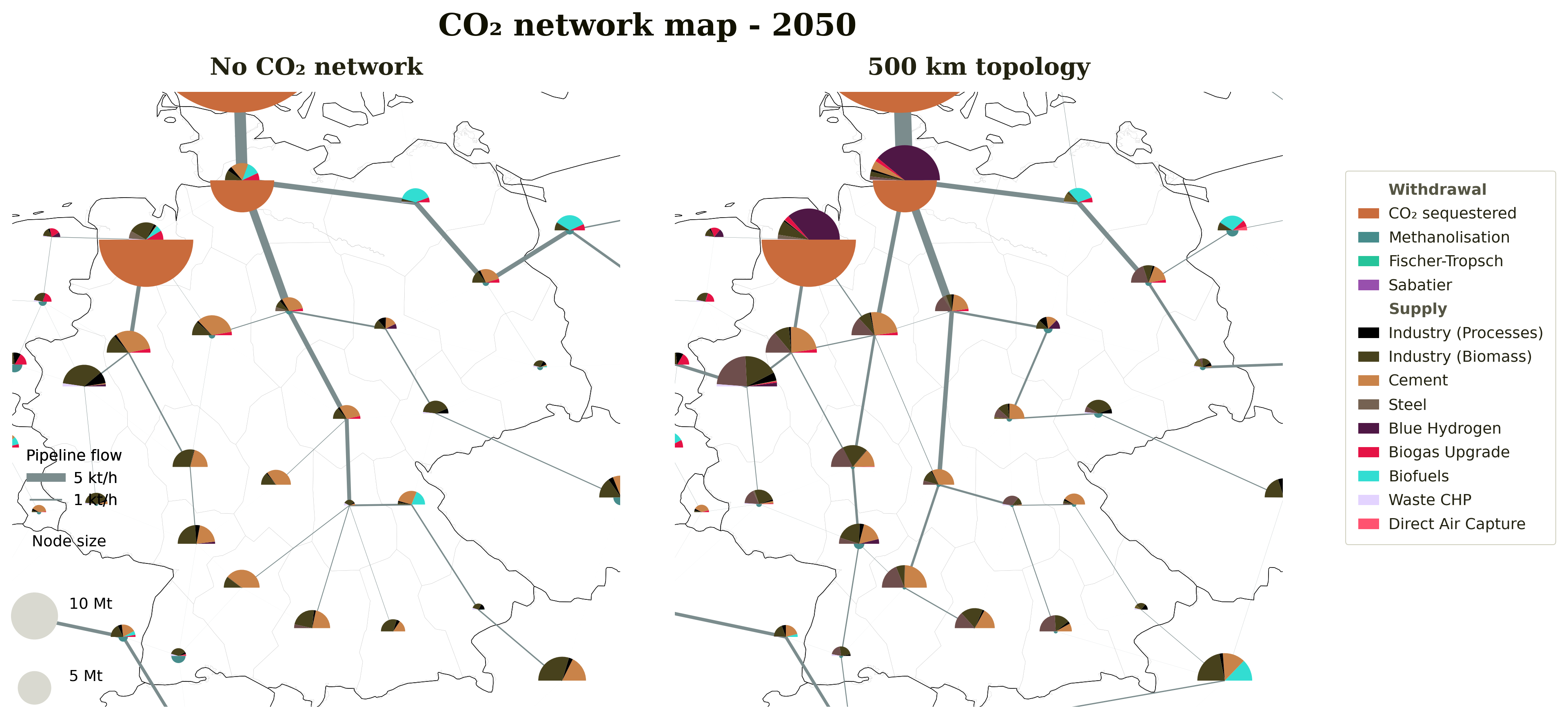}
    \caption{Carbon map for a scenario without carbon infrastructure in 2050 (left) and a \SI{500}{km} network (topology 002). Lines represent \ce{CO2} pipelines, pie charts represent \ce{CO2} captured (upper half) and utilized/sequestered (lower half).}
    \label{comp:co2_map_2050}
\end{figure}

\noindent Fig.~\ref{comp:co2_map_2035} and Fig.~\ref{comp:co2_map_2050} show the expansion of the \ce{CO2} network and volumes captured and utilized/sequestered for 2035 and 2050.
In the \textit{No \ce{CO2} network 2035} scenario (left), only small \ce{CO2} volumes are captured in the northern German regions and transported by ship to the \textit{Northern Lights} project in Norway.
By contrast, the topology scenarios (right) concentrate the majority of captured \ce{CO2} along the transport corridor extending eastwards from the Dutch interconnector through North Rhine-Westphalia and towards additional industrial source regions.
This highlights the importance of access to pipeline-based sequestration routes.
Without domestic transport infrastructure, carbon capture remains largely confined to regions with direct access to shipping, whereas the pipeline network enables the utilization of large industrial point sources located further inland.

In regions without access to \ce{CO2} transport infrastructure, including the \textit{No \ce{CO2} network 2035} scenario, captured carbon is utilized locally in methanol synthesis.
Hydrogen is transported to the location of carbon utilization and combined with locally available \ce{CO2}.

The 2050 network topology highlights one of the limitations of continuous infrastructure expansion within energy system models.
The resulting network contains a large number of small pipeline segments that would likely not be constructed in reality and do not necessarily correspond to practical infrastructure layouts.
For this reason, the main analysis focuses on aggregate infrastructure requirements measured in Mt/h\,km rather than on the detailed topology of the optimized 2050 network.

Next to CCS of domestic \ce{CO2} volumes, Germany becomes a transit country for European \ce{CO2} flows in 2050.
Captured \ce{CO2} from Poland, the Czech Republic, Austria and Switzerland is transported through Germany towards the available sequestration sites, reinforcing the importance of cross-border infrastructure coordination.

\noindent\textbf{Key model figures}

\noindent Table~\ref{tab:comparison} compares key indicators of topology~002, which connects North Rhine-Westphalia to the Netherlands with a maximum network length of \SI{500}{km}, against the \textit{No \ce{CO2} network 2035} scenario.

In 2035, the availability of \ce{CO2} transport infrastructure reduces both total European system costs and German consumer costs.
Total system costs decrease by \SI{8.3}{bnEUR/a}, while German consumer costs are reduced by \SI{22.1}{bnEUR/a}.
The larger reduction in consumer costs reflects the effect of lower scarcity prices throughout the German energy system.
Since consumer costs are calculated from marginal prices and demand, they are more sensitive to changes in system-wide scarcity than total system costs.
The carbon price decreases by approximately \SI{91}{EUR/tCO2} when \ce{CO2} transport infrastructure is available.
This indicates that the energy system can exploit comparatively inexpensive mitigation options by capturing industrial point sources and transporting the resulting \ce{CO2} to utilization and sequestration sites.
In the absence of transport infrastructure, the system must rely more heavily on alternative mitigation pathways such as blue hydrogen production, biofuel synthesis and biomass-based carbon dioxide removal.

The picture changes when the resulting systems are evaluated in 2050.
Under climate-neutrality constraints, the scenario without \ce{CO2} infrastructure in 2035 achieves lower total system costs and lower German consumer expenditures.
This result reflects a limitation of the myopic optimization framework.
Since the 2035 optimization does not anticipate the stricter climate constraints of 2050, the availability of \ce{CO2} transport infrastructure allows the system to postpone part of the decarbonization effort.
By contrast, the \textit{No \ce{CO2} network 2035} scenario is forced to electrify and decarbonize more aggressively already in 2035, which proves advantageous under the 2050 climate-neutrality constraint.

These findings suggest that the long-term value of early \ce{CO2} infrastructure deployment cannot be assessed conclusively within a myopic framework.
A perfect-foresight optimization would be required to determine whether the higher 2035 investments enabled by \ce{CO2} infrastructure ultimately reduce or increase total transition costs across the entire pathway to climate neutrality.

\begin{table}[htbp]
    \centering
    \caption{Comparison of topology~002 (\SI{500}{km}, NL sink) and the
    \textit{No \ce{CO2} network 2035} scenario.
    Positive $\Delta$ values indicate that the \textit{No \ce{CO2} network 2035} scenario is more expensive.}
    \label{tab:comparison}
    \setlength{\tabcolsep}{10pt}
    \renewcommand{\arraystretch}{1.25}
    \begin{tabular}{l S[table-format=4.2] S[table-format=4.2] S[table-format=+3.2]}
    \toprule
    \textbf{Metric}
        & {\textbf{No \ce{CO2} network}}
        & {\textbf{Topology 002}}
        & {$\boldsymbol{\Delta}$} \\
    \midrule
    \multicolumn{4}{l}{\textsc{2035}} \\[2pt]
    \quad TOTEX [bn\,EUR/a]
        & 938.4  & 930.1  & +8.3   \\
    \quad DE consumer costs [bn\,EUR/a]
        & 322.9  & 300.8  & +22.1  \\
    \quad Carbon price [EUR/t\ce{CO2}]
        & 482.48 & 391.95 & +90.53 \\[4pt]
    \multicolumn{4}{l}{\textsc{2050}} \\[2pt]
    \quad TOTEX [bn\,EUR/a]
        & 887.1  & 891.5  & -2.5   \\
    \quad DE consumer costs [bn\,EUR/a]
        & 237.0  & 257.0  & -20.0  \\
    \quad Carbon price [EUR/t\ce{CO2}]
        & 351.30 & 356.41 & -5.11 \\
    \bottomrule
    \end{tabular}
\end{table}

\section{Sensitivities}\label{app:sensitivities}

Our modeling results depend on a small number of constraints and modeling choices.
We therefore include two sensitivities: Section~\ref{sens:high_seq} presents results under a higher sequestration budget in 2035 and Section~\ref{sens:de_north_sea} explores a scenario in which Germany utilizes its domestic sequestration sites in the North Sea in 2035 already.

\subsection{Higher Sequestration Potential}\label{sens:high_seq}

\begin{figure}[ht]
    \centering
    \includegraphics[width=0.6\linewidth]{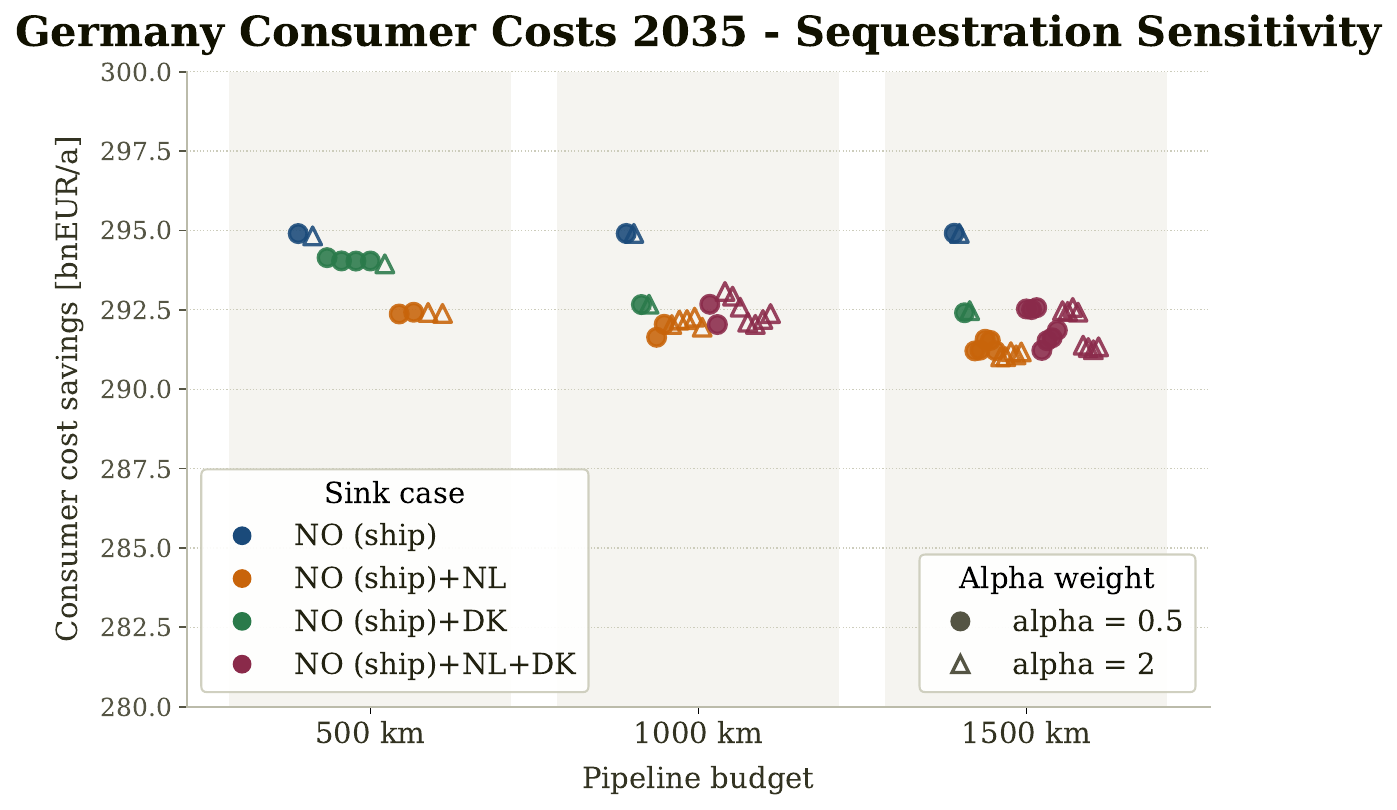}
    \caption{German consumer costs in 2035 under sequestration sensitivity. Each symbol represents one topology run and is grouped for different pipeline length limits. The color indicates the available sinks and the symbol the $\alpha$ weighting the availability of \ce{CO2} point sources for the development of the topology.}
    \label{sens:seq_cost}
\end{figure}

\begin{figure}[htbp]
    \centering
    \includegraphics[width=1.0\linewidth]{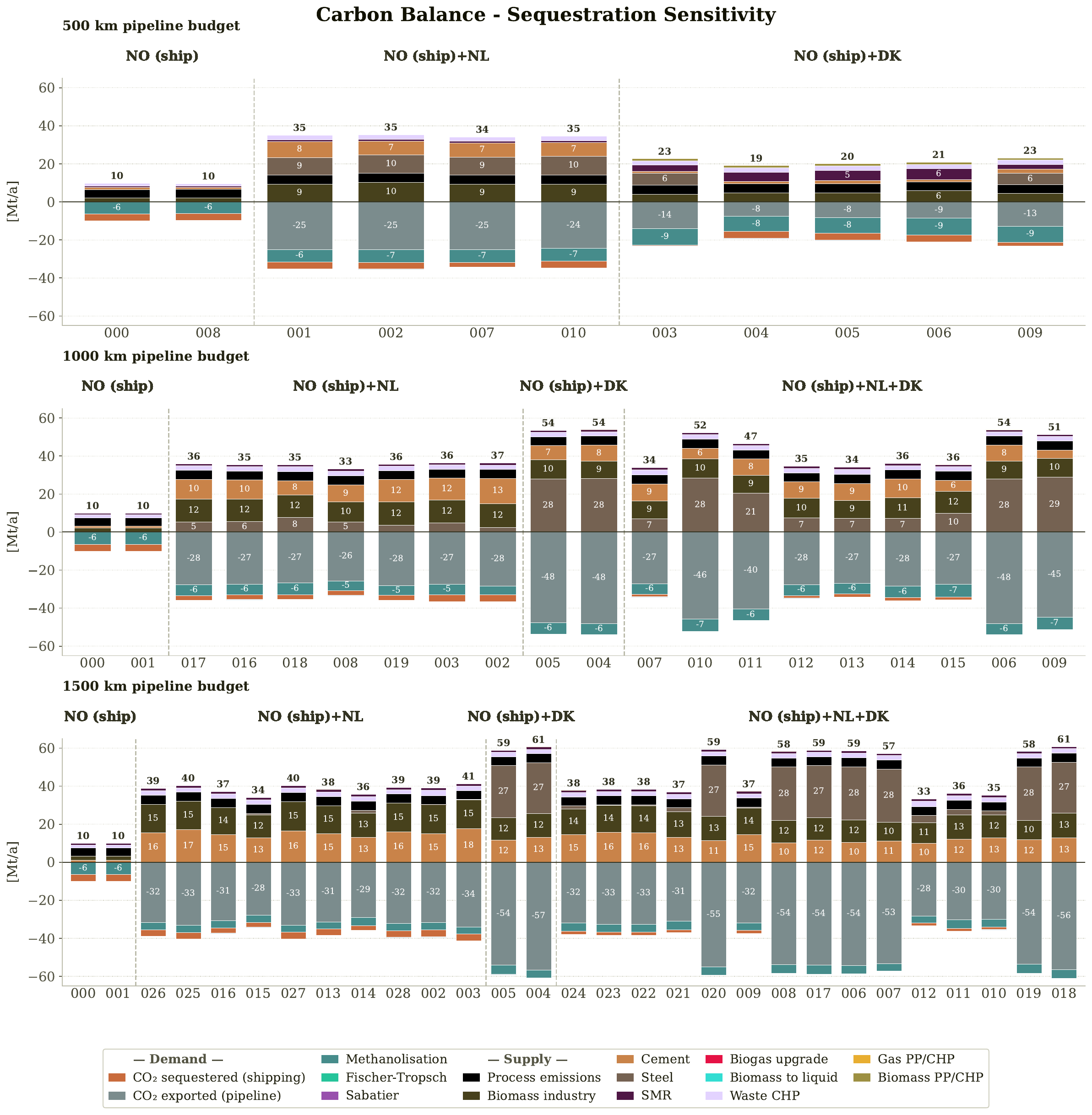}
    \caption{Carbon balances for different network lengths and sink cases in 2035 under sequestration sensitivity.}
    \label{sens:seq_balance}
\end{figure}

\begin{figure}[htbp]
    \centering
    \includegraphics[width=1.0\linewidth]{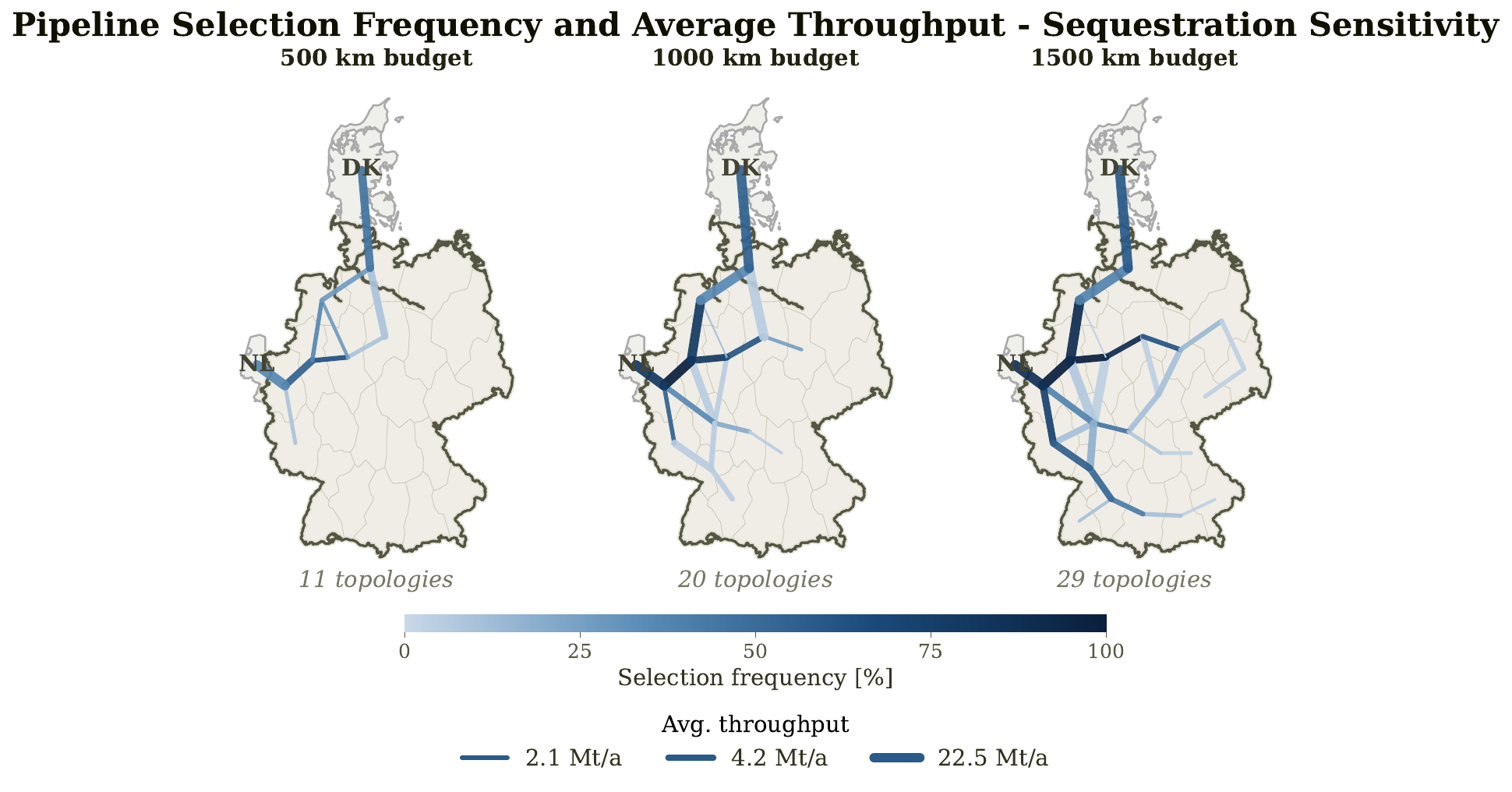}
    \caption{Selction frequency of pipeline segments from graph theory (color) and if chosen average throughput of CO2 (thickness) in 2035 under sequestration sensitivity.}
    \label{sens:seq_map}
\end{figure}

Increasing the available sequestration potential in 2035 from \SI{100}{Mt/a} to \SI{200}{Mt/a} amplifies the role of \ce{CO2} transport infrastructure (Appendix~\ref{app:sensitivities}).
In contrast to the results presented in the main part of the paper, German consumer costs become more sensitive to network topology and sink accessibility (Fig.~\ref{sens:seq_cost}).
Under a tight pipeline budget of \SI{500}{km}, access to the Dutch interconnector is sufficient to capture most of the economic benefit, whereas scenarios relying exclusively on shipping to Norway remain more expensive.
The larger storage potential also leads to higher overall capture and sequestration volumes (Fig.~\ref{sens:seq_balance}).
While scenarios with only shipping to the \textit{Northern Lights} project remain limited to approximately \SI{10}{MtCO2/a} of captured \ce{CO2}, the expanded storage capacity shifts the capture merit order towards additional industrial sources, including steel production.
Consequently, pipeline utilization increases across the network, with the major transport corridors carrying substantially larger \ce{CO2} volumes (Fig.~\ref{sens:seq_map}).

\newpage
\subsection{Early German North Sea Sequestration}\label{sens:de_north_sea}

\begin{figure}[htbp]
    \centering
    \includegraphics[width=0.6\linewidth]{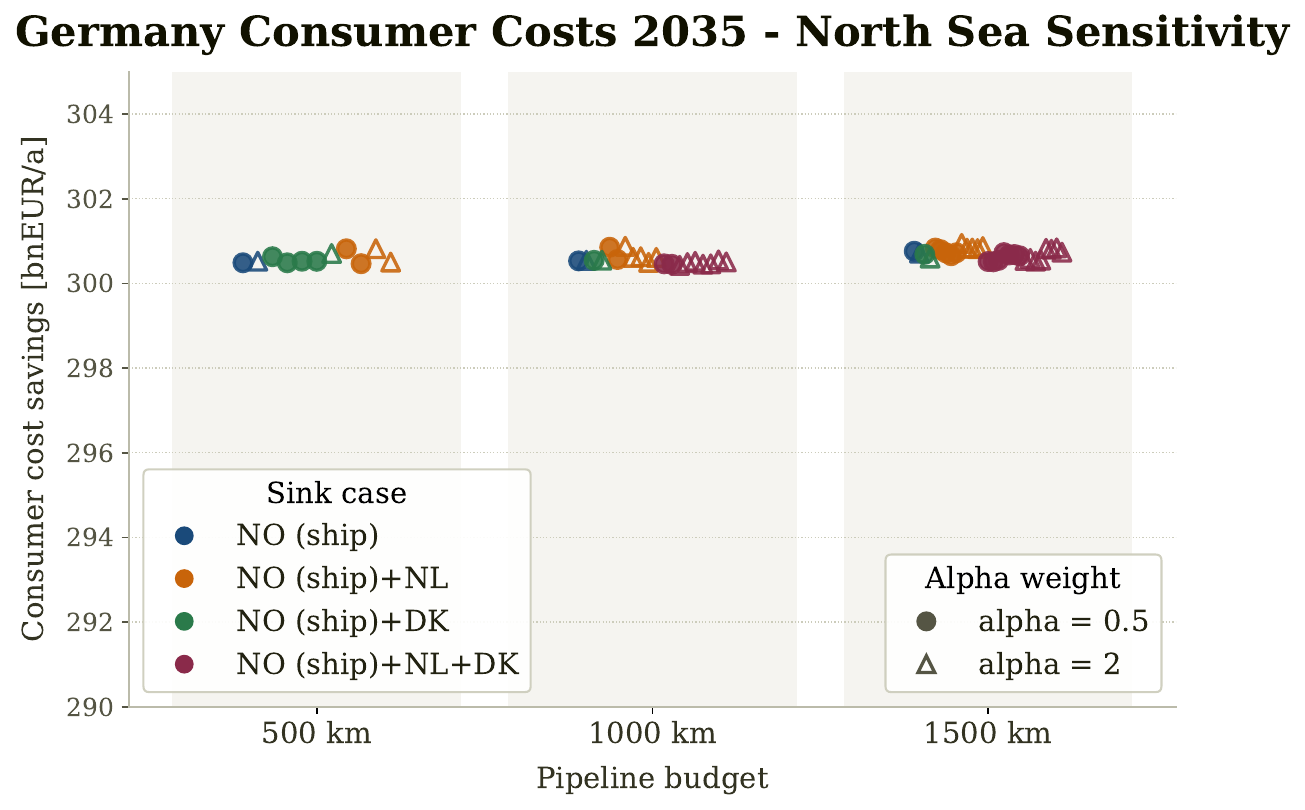}
    \caption{German consumer costs in 2035 under availability of German North Sea sensitivity. Each symbol represents one topology run and is grouped for different pipeline length limits. The color indicates the available sinks and the symbol the $\alpha$ weighting the availability of \ce{CO2} point sources for the development of the topology.}
    \label{sens:ns_cost}
\end{figure}

\begin{figure}[htbp]
    \centering
    \includegraphics[width=1.0\linewidth]{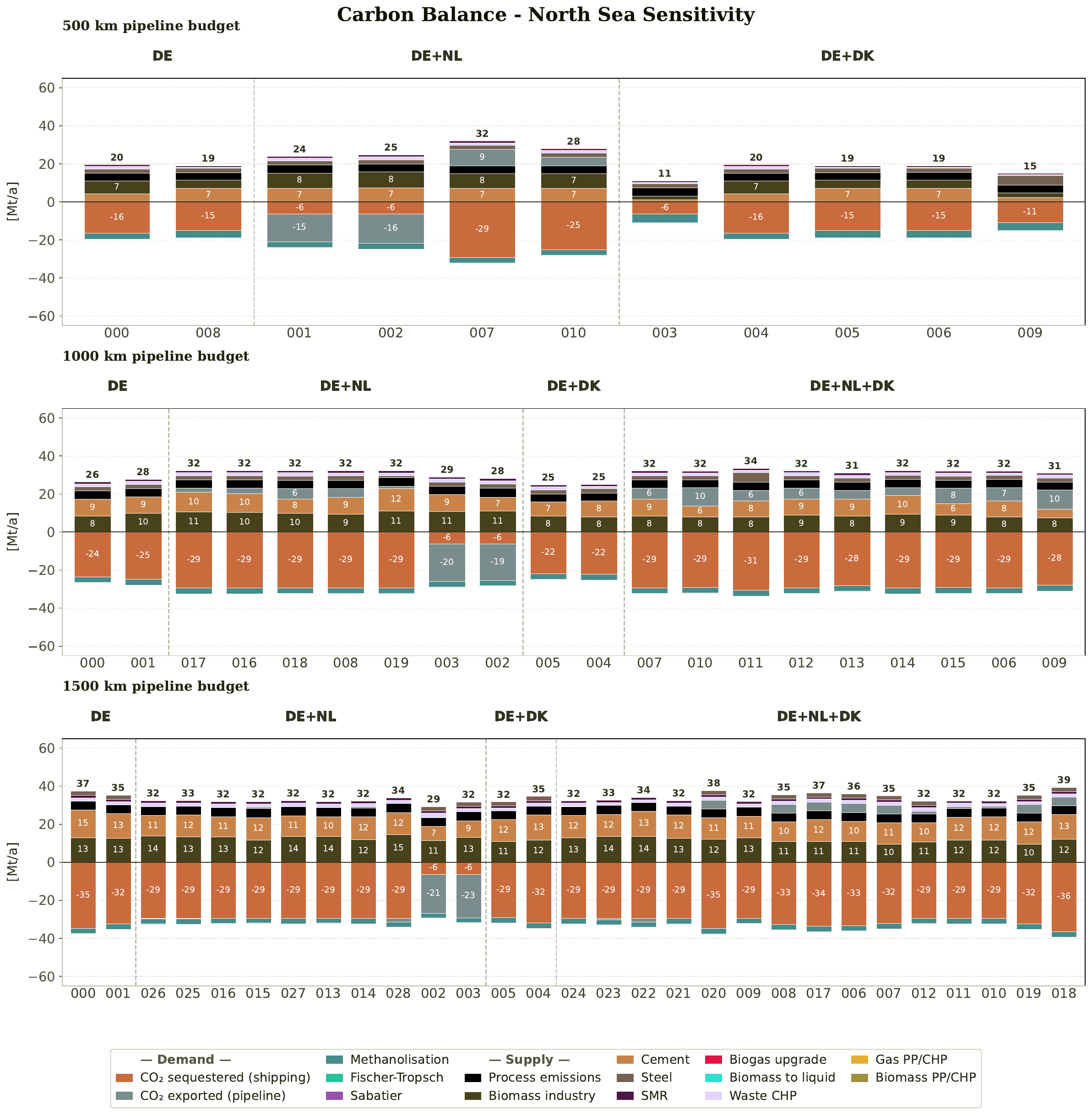}
    \caption{Carbon balances for different network lengths and sink cases in 2035 under availability of German North Sea.}
    \label{sens:ns_balance}
\end{figure}

\begin{figure}[htbp]
    \centering
    \includegraphics[width=1.0\linewidth]{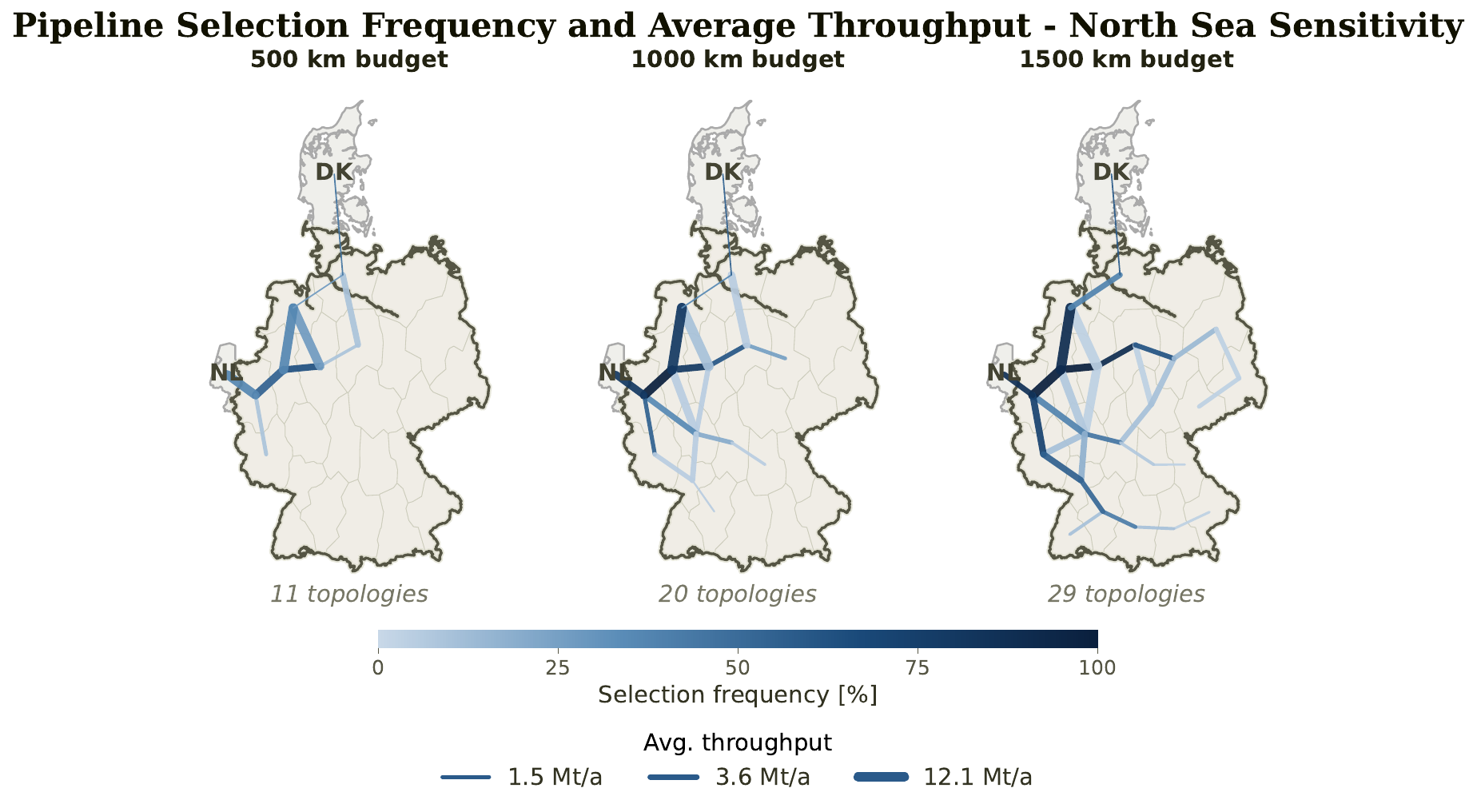}
    \caption{Selction frequency of pipeline segments from graph theory (color) and if chosen average throughput of CO2 (thickness) in 2035 under availability of German North Sea sensitivity.}
    \label{sens:ns_map}
\end{figure}

Allowing permanent sequestration in the German North Sea already in 2035 while keeping the overall sequestration potential fixed at \SI{100}{MtCO2/a} has only a minor effect on German consumer costs (Fig.~\ref{sens:ns_cost}).
The overall amount of captured \ce{CO2} also remains largely unchanged (Fig.~\ref{sens:ns_balance}).
However, the spatial allocation of sequestration changes considerably.
If no international sink is available, domestic offshore storage substantially increases capture and sequestration compared to shipping to the \textit{Northern Lights} project.
More generally, the availability of the German North Sea reduces the reliance on \ce{CO2} exports via international interconnectors, as most captured \ce{CO2} is permanently stored domestically.
In several scenarios, Germany even becomes a net importer of \ce{CO2}, providing sequestration capacity for neighbouring countries.
Consequently, the overall pipeline utilization remains similar to the base case, although transport volumes shift towards the German North Sea storage region (Fig.~\ref{sens:ns_map}).

\end{document}